\documentclass[preprint]{article}
\RequirePackage[OT1]{fontenc}

\usepackage{amsthm,amsmath,amssymb,mathrsfs,graphicx}
\usepackage{verbatim,multirow,subfigure}
\usepackage{rotating, booktabs}
\usepackage{enumerate,color}
\usepackage{array}
\RequirePackage{hyperref}
\usepackage[margin = 1in]{geometry}
\usepackage{setspace}
\usepackage{authblk}
\usepackage[ruled,vlined]{algorithm2e}

\usepackage{natbib}
\usepackage{soul}
\usepackage{array}
\newcolumntype{L}[1]{>{\raggedright\let\newline\\\arraybackslash\hspace{0pt}}m{#1}}
\newcolumntype{C}[1]{>{\centering\let\newline\\\arraybackslash\hspace{0pt}}m{#1}}
\newcolumntype{R}[1]{>{\raggedleft\let\newline\\\arraybackslash\hspace{0pt}}m{#1}}

\theoremstyle{plain}

\newtheorem{theorem}{Theorem}[section]

\newcommand{\tr}{{\rm tr}}

\newcommand{\muvec}{\boldsymbol{\mu}}
\newcommand{\xvec}{\mathbf{X}}
\newcommand{\yvec}{\mathbf{Y}}
\newcommand{\zvec}{\mathbf{Z}}
\newcommand{\xvecstar}{\mathbf{X}^*}
\newcommand{\yvecstar}{\mathbf{Y}^*}

\allowdisplaybreaks

\title{Covariance matrix testing in high dimension using random projections}

\author[1]{Deepak Nag Ayyala \thanks{Corresponding author: 1120 15th Street, AE 1040, Augusta, GA 30912; dayyala@augusta.edu}}
\author[1]{Santu Ghosh}
\author[1]{Daniel F. Linder}
\affil[1]{Department of Population Health Sciences, Medical College of Georgia, Augusta University}
\date{}

\begin{document}

\maketitle
\begin{abstract}
	Estimation and hypothesis tests for the covariance matrix in high dimensions is a challenging problem as the traditional multivariate asymptotic theory is no longer valid. When the dimension is larger than or increasing with the sample size, standard likelihood based tests for the covariance matrix have poor performance. Existing high dimensional tests are either computationally expensive or have very weak control of type I error. In this paper, we propose a test procedure, \textbf{CRAMP}, for testing hypotheses involving one or more covariance matrices using random projections. Projecting the high dimensional data randomly into lower dimensional subspaces  alleviates of the \textit{curse} of dimensionality, allowing for the use of traditional multivariate tests. An extensive simulation study is performed to compare CRAMP against asymptotics-based high dimensional test procedures. An application of the proposed method to two gene expression data sets is presented.

\end{abstract}

%\begin{keyword}
%	high dimension; covariance matrix; hypothesis testing; random projections
%\end{keyword}
\section{Introduction}

In multivariate analysis, hypothesis tests involving the first two moments - mean and covariance matrix have been extensively studied. Consider a random variable $\xvec$ with mean $\muvec = \mathbb{E}(\xvec)$ and covariance matrix $\Sigma = \mathbb{E} \left\{ (\xvec - \muvec) (\xvec - \muvec)^{\top} \right\}$. There is a vast amount of literature for inference involving the mean $\muvec$, starting with the Hotelling's $T^2$ statistic. Refer to \cite{Ayyala2020} for an extensive review of methods for the mean vector testing. In this article, we focus on inference on the covariance matrix. Given a random sample from a $p$-dimensional Gaussian distribution with mean $\muvec$ and variance $\Sigma$, we are interested in testing the hypothesis 
\begin{equation}
H_0 : \Sigma = \Sigma_0 \hspace{1cm} \mbox{ vs. } \hspace{1cm} H_A : \Sigma \neq \Sigma_0,
\label{eqn:onesampleintro}
\end{equation}
for some known $p \times p$ matrix $\Sigma_0$. Of specific interest is when $\Sigma_0$ corresponds to a particular geometric shape - $\Sigma_0 = \sigma^2 \mathcal{I}_p, \sigma \in \mathbb{R}$ for a {\it spherical} normal distribution or $\Sigma_0 = {\rm diag}(\sigma_{1}, \ldots, \sigma_{p})$ for {\it independent} components. Other forms, such as block diagonal can be helpful in testing the presence of groups of independent elements in $\xvec$. In the two sample case, it is of interest to compare the covariance matrices $\Sigma_1$ and $\Sigma_2$ of two populations $\xvec$ and $\yvec$ respectively. Equality of covariance matrices implies the distributions of $\xvec$ and $\yvec$ have the same shape, but are centered at different locations.  Importance of the problem of testing equality of covariance matrices for Gaussian models lies in the network interpretation of the covariance matrix. The inverse of the covariance matrix, called the {\it precision matrix} is used to construct undirected graphical network models with elements of the variable as nodes \cite{Zhao2014, Cai2012}. %\DNA{Two more sentences about the importance of comparing two covariance matrices}

For both the one and two sample hypotheses, traditional likelihood ratio tests are developed and studied in great detail \cite{Anderson2003}. However the tests are valid only when $p < n$ and $p$ is fixed. For high dimensional data, i.e. when $p$ increases with $n$ or when $p > n$, the asymptotic properties of these tests are no longer valid. This is because the sample covariance matrix $\mathcal{S}$ has rank $\min(p, n - 1)$, where $n$ is the sample size. Therefore unconstrained estimation will lead to rank-deficient and inconsistent estimators when $p > n$. To avoid this problem, test statistics can be constructed based on a real-valued functional of $\mathcal{S}$. This approach is very commonly used in high dimensional inference for the mean \cite{Ayyala2020}. For example in the one-sample hypothesis in \eqref{eqn:onesampleintro}, we can use $f(\Sigma) = \tr \left( \Sigma - \Sigma_0 \right)^2$ as the functional, with the rejection region determined by studying the asymptotic properties of the sampling distribution of $f(\mathcal{S})$.  Appropriate functionals for the two-sample case can be constructed similarly.

An approach that is gaining prominence in other domains of high dimensional inference but has not been implemented explicitly in covariance matrix testing is the use of random projections. A computationally intensive approach, a random projection (RP) based inference involves embedding the original $p$-dimensional data into a lower $k$-dimensional space using linear projections. Dimension of the embedded space $k$ can be chosen to be smaller than $n$, thereby upholding the assumptions of traditional multivariate methods. Validity of this method is guaranteed by the Johnson-Lindenstrauss lemma \cite{Johnson1984}. Random projection methods have been used for the mean vector test \cite{Lopes2011, Srivastava2014}, ensemble machine learning methods such as classification \citep{Cannings2017}. To the best of our knowledge, this approach has not been used in hypothesis testing for the covariance matrix.

In this paper, we propose \textbf{CRAMP} - {\bf c}ovariance matrix testing using {\bf ra}ndom {\bf m}atrix {\bf p}rojections. The rest of the article is organized as follows. In section \ref{sec:hypotheses}, we introduce two specific one sample tests and the two sample test hypotheses. A literature review of existing test procedures in both traditional and high-dimensional settings is also provided. Random projection based tests is introduced in section \ref{sec:proposedtest}. Theoretical details and algorithms for the one and two sample tests are also explicitly described. In section \ref{sec:simulation}, an extensive simulation study comparing the different methods is presented. We applied CRAMP to test equivalence of gene networks, which are represented by the covariance matrices using gene expression data. Results from the analysis of these data sets are presented in section \ref{sec:realdata}.

\section{Hypotheses for covariance matrices}
\label{sec:hypotheses}
\subsection{One sample tests}
Consider a random sample $\xvec_1, \ldots, \xvec_n$ from a $p$-dimensional continuous distribution $\mathcal{F}_p$ with mean $\muvec$ and variance $\Sigma$. The parameter of interest for this study is $\Sigma$, the variance matrix. We are interested in testing the hypotheses
\begin{gather}
H_{0I} : \Sigma = \mathcal{I}_p \hspace{5mm} \mbox{ vs. } \hspace{5mm} H_{1I} : \Sigma \neq \mathcal{I}_p, \label{eqn:uniformity}\\
H_{0S} : \Sigma = \sigma^2 \mathcal{I}_p \hspace{5mm} \mbox{ vs. } \hspace{5mm} H_{1S} : \Sigma \neq \sigma^2 \mathcal{I}_p, \label{eqn:sphericity}
\end{gather}
where $\mathcal{I}_p$ is the identity matrix of dimension $p$ and $\sigma > 0$ is an unknown parameter. The hypotheses in \ref{eqn:uniformity} and \ref{eqn:sphericity} are commonly referred to as tests for {\it identity} and {\it sphericity} respectively. The general test for $H_0 : \Sigma = \Sigma_0$ for some known matrix $\Sigma_0$ can be viewed as a test for identity when the data is transformed as $\xvec \mapsto \Sigma_0^{-1/2} \xvec$. The hypotheses can be equivalently stated in terms of the eigenvalues of $\Sigma$. If $\lambda_1, \ldots, \lambda_p$ denote the eigenvalues of $\Sigma$, then \ref{eqn:uniformity} and \ref{eqn:sphericity} can be stated as
\begin{gather*}
H_{0I} : \lambda_i = 1 \,\,\, \forall \,\,\, i \hspace{5mm} \mbox{ vs. } \hspace{5mm} H_{1I} : \lambda_i \leq 1 \mbox{  for at least one } i, \label{eqn:uniformity2} \\
H_{0S} : \lambda_1 = \ldots = \lambda_p \hspace{5mm} \mbox{ vs. } \hspace{5mm} H_{1S} : \lambda_i \neq \lambda_j \mbox{ for some } i \neq j. \label{eqn:sphericity2}
\end{gather*}

Let $\mathcal{S} = n^{-1} \sum\limits_{i = 1}^n \left(\xvec_i - \overline{\xvec} \right) \left( \xvec_i - \overline{\xvec} \right)^{\top}$ denote the sample covariance matrix, where $\overline{\xvec} = n^{-1} \sum\limits_{i=1}^n \xvec_i$ is the sample mean. When $\mathcal{F}_p$ is the Gaussian distribution, $\mathcal{S}$ is the maximum likelihood estimator which follows a Wishart distribution. The likelihood ratio tests for the two tests are given by
\begin{gather}
	\begin{split}
		LRT_{I} = (n - 1) \left\{ 1 - \frac{1}{6n - 7} \left(2p + 1 - \frac{2}{p + 1} \right)  \right\} \left[ -\log(|\mathcal{S}|) + \tr(\mathcal{S}) - p \right], \\
		LRT_{S} = - \left\{ n - 1 - \frac{2p^2 + p + 2}{6p}  \right\} \left[ p \log p + \sum\limits_{i = 1}^p \log \lambda_i - p \log \left( \sum\limits_{i  = 1}^p \lambda_i \right) \right].
	\end{split}
\end{gather}
Under the null hypothesis, the test statistics are approximately distributed as a $\chi^2$ distribution with degrees of freedom $\nu = p(p + 1)/2$ and $\nu = p(p + 1)/2 - 1$ respectively \cite{Rencher}.

Another approach to test the hypotheses is to construct a functional of the covariance matrix which will be zero under the null hypothesis and non-zero under the alternative. For sphericity and identity, it is straightforward to see that the functionals 
\begin{equation*}
\mathcal{U} = \frac{1}{p} \tr \left\{  \frac{\Sigma}{\tr \Sigma/p} - \mathcal{I}_p \right\}^2, \hspace{1cm} 
\mathcal{V} = \frac{1}{p} \tr \left\{ \Sigma - \mathcal{I}_p \right\}^2,
\end{equation*}
are non-negative and are equal to zero under $H_{0S}$ and $H_{0I}$ respectively. Using these functionals, \citep{John1972} and \citep{Nagao1973} proposed the following test statistics by plugging in the sample covariance matrix estimate to test $H_{0S}$ and $H_{0I}$ respectively:
\begin{equation}
U_{John} = \frac{1}{p} \tr \left\{ \frac{ \mathcal{S}}{\tr \mathcal{S}/p}  - \mathcal{I}_p \right\}^2, \hspace{1cm} V_{Nagao} = \frac{1}{p} \tr \left\{ \mathcal{S} - \mathcal{I}_p \right\}^2.
\label{eqn:johnnagao}
\end{equation}
It is shown that under the null hypothesis, $U_{John}$ and $V_{Nagao}$ are asymptotically distributed as chi-squared random variables with $p(p + 1)/2 - 1$ degrees of freedom. When the sample size is small, \citep{Nagao1973} also provided second-order corrections to the p-values for both test statistics. While these tests are constructed assuming normality of the samples, they are applicable even when $\mathcal{S}$ is singular, unlike the likelihood ratio tests which involves inverting the sample covariance matrix. However, these tests fail when the data is high-dimensional, i.e. when $p$ is larger than $n$. While the tests can be applied in practice, the asymptotic properties fail to hold unless $p$ is assumed to be fixed with respect to $n$.

Under high dimensional setting, \citep{Ledoit2002} studied the properties of $U_{John}$ and $V_{Nagao}$ for high-dimensional models when $p/n \rightarrow c \in (0, \infty)$. They observed that $U_{John}$ is consistent for high-dimensional data, whereas $V_{Nagao}$ fails when $p$ increases with $n$. Modifying $V_{Nagao}$, they proposed 
\begin{equation}
V_{LW} = \frac{1}{p} \tr \left\{ \mathcal{S} - \mathcal{I}_p \right\}^2 - \frac{p}{n} \left\{ \frac{ \tr \mathcal{S} }{p} \right\}^2 + \frac{p}{n}.
\end{equation}
Under $H_{0I}$, $V_{LW}$ is shown to asymptotically follow a $\chi^2$ distribution with $p(p + 1)/2$ degrees of freedom. The asymptotic distribution is derived under a normal model for the observations. 

With increased interest in high dimensional inference, several other tests have been proposed for the hypotheses in \ref{eqn:uniformity} and \ref{eqn:sphericity}. \citep{Srivastava2014} proposed using modified estimators of $\tr \Sigma$ and $\tr \Sigma^2$ in $\mathcal{U}$ and $\mathcal{V}$. Their test statistic is given by
\begin{equation}
U_{SYK} = \frac{n - 1}{2} \left[ \frac{\widehat{a}_2}{\widehat{a}_1} - 1 \right], \hspace{1cm} V_{SYK} = \frac{n - 1}{2} \left[ \widehat{a}_2 - 2 \widehat{a}_1 + 1 \right],
\label{eqn:srivastava1}
\end{equation}
where $\widehat{a}_1 = \tr \left(\mathcal{S} \right)/p$, $\widehat{a}_2 = \left\{ pn(n - 1)(n - 2)(n - 3)  \right\}^{-1} \bigg[ (n - 1)^3(n - 2) \tr \mathcal{S}^2 - n(n - 1)^3 \tr \left(\mathcal{D}_{\mathcal{S}}^2 \right) + (n - 1)^2 \tr \left( \mathcal{S}^2 \right) \bigg]$ and $\mathcal{D}_{\mathcal{S}}$ denotes the diagonal of the sample covariance matrix. The test statistics are shown to be asymptotically normally distributed under $H_{0S}$ and $H_{0I}$ respectively. The statistics in \eqref{eqn:srivastava1} are based on comparing the arithmetic means of the eigenvalues of $\Sigma^k$ for $k = 1, 2$. Extending the result to higher order powers, \citep{Fisher2010, Fisher2012} expanded it to the fourth powers of $\Sigma$ and \citep{Qian2020} extended the results to the sixth power. \citep{Chen2010} used Hoeffding's $U$-statistics to estimate $\tr \Sigma$ and $\tr \Sigma^2$. Their test statistics are given by
\begin{equation}
U_{CZZ} = p \left( \frac{T_{2,n}}{T_{1,n}^2} \right) - 1, \hspace{1cm} V_{CZZ} = \frac{1}{p}T_{2,n} - \frac{2}{p} T_{1,n} + 1,
\label{eqn:chen1}
\end{equation}
where $T_{1,n} = n^{-1} \sum_{i=1}^n \xvec_i^{\top} \xvec_i -  \{ n(n-1)\}^{-1} \sum_{i \neq j} \xvec_i^{\top} \xvec_j$ is the $U$-estimator for $\tr \Sigma$ 
and 
\begin{equation*}
T_{2,n} = \frac{ \sum\limits_{i \neq j} \left(\xvec_i^{\top} \xvec_j \right)^2}{n(n-1)} - \frac{2 \sum\limits_{i \neq j \neq k} \xvec_i^{\top} \xvec_j \xvec_j^{\top} \xvec_k }{n(n-1)(n-2)}  + \frac{\sum\limits_{i \neq j\neq k \neq l} \xvec_i^{\top} \xvec_j \xvec_k^{\top} \xvec_l}{n(n-1)(n-2)(n-3)} 
\end{equation*}
is the $U$-estimator for $\tr \Sigma^2$. Under the null hypotheses, the test statistics $nU_{CZZ}/2$ and $nV_{CZZ}/2$ both asymptotically follow a standard normal distribution. 

\subsection{Two sample tests}

In the two sample case, our interest lies in comparing the covariance matrices of two independent populations. Let $\xvec_1, \ldots, \xvec_n$ and $\yvec_1, \ldots, \yvec_m$ be random samples drawn from $p$-dimensional distributions $\mathcal{F}_p$ and $\mathcal{G}_p$ respectively. Denoting the covariances of the two populations by $\Sigma_1$ and $\Sigma_2$ respectively, the hypothesis of interest is 
\begin{equation}
H_{0T} : \Sigma_1 = \Sigma_2 \hspace{5mm} \mbox{ vs. } \hspace{5mm} H_{1T} : \Sigma_1 \neq \Sigma_2. 
\label{eqn:twosample}
\end{equation}
Let $\mathcal{S}_1 =  n^{-1} \sum\limits_{i = 1}^n \left(\xvec_i - \overline{\xvec} \right) \left( \xvec_i - \overline{\xvec} \right)^{\top}$ and $\mathcal{S}_2 =  m^{-1} \sum\limits_{i = 1}^m \left(\yvec_i - \overline{\yvec} \right) \left( \yvec_i - \overline{\yvec} \right)^{\top}$ denote the sample covariance matrices of the two populations respectively. Let $\mathcal{S}_{pl} = (n \mathcal{S}_1 + m \mathcal{S}_2)/(n + m)$ denote the pooled sample covariance matrix. When $p < \min(m, n)$ and both $\mathcal{F}_p$ and $\mathcal{G}_p$ are assumed to be Gaussian, the likelihood ratio test is constructed using
\begin{equation}
\mathcal{M} = \frac{ \left| \mathcal{S}_1 \right|^{n - 1} \left| \mathcal{S}_2\right|^{m - 1}  }{ \left| \mathcal{S}_{pl} \right|^{n + m - 2} }.
\label{eqn:lrt}
\end{equation}
Under $H_{0T}$, $T = -2 (1 - c_1) \mathcal{M}$ is asymptotically $\chi^2$-distributed with $p(p + 1)/2$ degrees of freedom, where $c_1 = (1/n + 1/m - 1/(n + m)) \frac{2p^2 + 3p - 1}{6(p + 1)}$. This test, called the Box's $\mathcal{M}$-test, also has an approximation yielding an $F$ distribution in the limit. For lower dimensional models ($p < n$), a Wald-type test can also be constructed as
\begin{equation}
T_{Wald} = \frac{n + m}{2} \left[ \frac{n}{n + m} \tr \left( \mathcal{S}_1 \mathcal{S}_{pl} \right)^2 + \frac{m}{n + m} \tr \left( \mathcal{S}_2 \mathcal{S}_{pl}^{-1} \right)^2 - \frac{ n m}{(n + m)^2} \tr \left( \mathcal{S}_1 \mathcal{S}_{pl}^{-1} \mathcal{S}_2 \mathcal{S}_{pl}^{-1} \right) \right],
\label{eqn:waldtest}
\end{equation}
which follows a $\chi^2$ distribution asymptotically with $p(p + 1)/2$ degrees of freedom under $H_{0T}$. 

However, the above two tests fail for high dimensional models with $p > n$. Similar to the one-sample tests, one way to avoid specifying a distribution model to the two groups is by constructing a functional of $\Sigma_1$ and $\Sigma_2$ which is zero under $H_{0T}$ and non-zero otherwise. The Wald-test in \ref{eqn:waldtest} can be thought of as being based on this principle with $\tr \left( \Sigma_1 \Sigma_2^{-1} \right)$  as the functional. However in high dimensional inference, sample covariance matrices are singular and hence matrix inversion is usually avoided. Instead, a more commonly used functional to compare covariance matrices is $\tr \left( \Sigma_1 - \Sigma_2 \right)^2$, which is equivalent to the Frobenius norm of the difference $\Sigma_1 - \Sigma_2$. 

When the samples are normally distributed, \citep{Schott2007} proposed a test statistic when $p/n \rightarrow b \in [0, \infty)$. Under the assumption that $\lim \tr (\Sigma_i^k)/p = b \in (0, \infty)$ for $i = 1, 2$ and $k = 1, \ldots, 8$, the test statistic
\begin{align}
T_{Sch} & = \tr \left( \mathcal{S}_1 - \mathcal{S}_2 \right)^2  - \frac{n - 2}{(n + 1)(n - 1)} \left\{ (n - 1)(n - 3) \tr \left(\mathcal{S}_1^2 \right) + (n - 1) \tr \left(\mathcal{S}_1 \right)^2 \right\} \nonumber \\
&  - \frac{m - 2}{(m + 1)(m - 1)} \left\{ (m - 1)(m - 3) \tr \left( \mathcal{S}_2^2 \right) + (m - 1)^2 \tr \left( \mathcal{S}_2 \right)^2 \right\} 
\label{eqn:schott}
\end{align}
is shown to be asymptotically normal under $H_{0T}$. This test statistic is still restrictive in terms of the distributional assumption required to derive the asumptotic properties. 

Relaxing the normality assumption, \citep{Srivastava2014} considered a factor linear model  of the form $\xvec = \muvec + \boldsymbol{F} \boldsymbol{u}$, for some $p \times m$ matrix $\boldsymbol{F}$ and $m \times 1$ random vector $\boldsymbol{u}$. The distributional assumption on $\xvec$ is replaced by conditions on the moments of elements of $\boldsymbol{u}$. The test statistic, which is constructed based on the function $\tr \left( \mathcal{S}_1 - \mathcal{S}_2 \right)^2$, is given by
\begin{gather}
T_{SYK} = \frac{ \Delta_1 + \Delta_2 - 2 p^{-1} \tr \left( \mathcal{S}_1 \mathcal{S}_2 \right) }{ 2 \left( \frac{1}{n - 1} + \frac{1}{m - 1} \right) \frac{ (n - 1) \Delta_1 + (m - 1) \Delta_2}{n + m - 2} },
\label{eqn:srivastava2}
\end{gather}
where $\Delta_k = \left\{ (n_k - 1)^3 (n_k - 2) \tr \left( \mathcal{S}_k^2 \right) - n_k (n_k - 1)^3 \tr \left( D_{\mathcal{S}_k}^2 \right) + (n_k - 1)^2 \tr \left( \mathcal{S}_k \right)^2  \right\}/ \{p n_k (n_k - 1)(n_k - 2)(n_k - 3) \}$ for k = 1, 2 with $n_1 = n$ and $n_2 = m$. The dimension is allowed to increase at a polynomial rate with respect to the sample size, $p = O(n^{\delta})$ for $1/2 < \delta < 1$. Under $H_{0T}$, the test statistic is shown to converge to a standard normal distribution.

Using $\tr \left( \Sigma_1 - \Sigma_2 \right)^2$ as the functional, \citep{Li2012} developed a test statistic. The main idea behind the test statistic is to use Hoeffding's $U$-statistics to construct unbiased estimators for the functional. Asymptotic properties of this estimator are used to develop the test procedure. The test statistic, given by
\begin{gather}
T_{LC} = \frac{ \mathcal{A}_{n,1} + \mathcal{A}_{m,2} - 2 \mathcal{C}_{n m}}{ \sigma_{n,m}},
\end{gather} 
where for $h = 1, 2$,
\begin{align*}
A_{n, h} & = \frac{1}{n(n - 1)} \sum\limits_{i \neq j} \left( \xvec_{hi}^{\top} \xvec_{hj} \right)^2 - \frac{2}{n (n - 1)(n - 2)} \sum\limits_{i \neq j \neq k} \xvec_{hi}^{\top} \xvec_{hj} \xvec_{hi}^{\top} \xvec_{hk} \\
& + \frac{1}{n(n - 1)(n - 2)(n - 3)} \sum\limits_{i \neq j \neq k \neq \ell} \xvec_{hi}^{\top} \xvec_{hj} \xvec_{hk}^{\top} \xvec_{h \ell},
\end{align*}
with $\xvec_{1i} = \xvec_{i}$ and $\xvec_{2i} = \yvec_{i}$ and 
\begin{align*}
C_{n, m} & = \frac{1}{nm} \sum\limits_{i = 1}^n \sum\limits_{j = 1}^m \left( \xvec_i^{\top} \yvec_j \right)^2 - \frac{1}{n (n - 1) m} \sum\limits_{i \neq j} \sum\limits_k \xvec_i^{\top} \yvec_k \xvec_j^{\top} \yvec_k \\
&  - \frac{1}{m ( m - 1) n}   \sum\limits_{i \neq j} \sum\limits_k \yvec_i^{\top} \xvec_k \yvec_j^{\top} \xvec_k + \frac{1}{n(n-1)m(m - 1)} \sum\limits_{i \neq k} \sum\limits_{j \neq \ell} \xvec_i^{\top} \yvec_j \xvec_k^{\top} \yvec_{\ell}.
\end{align*}
Under regularity conditions on the covariance matrices, $T_{LC}$ is asymptotically normal under $H_{0T}$. One of the main advantages of $T_{LC}$ over 
$T_{SYK}$ and $T_{Sch}$ is that a direct relationship between $n$ and $p$ has been relaxed. 

In the above two test statistics, the aggregate difference between $\Sigma_1$ and $\Sigma_2$ is measured using the Frobenius norm. \citep{Cai2013} proposed a test based on the maximum difference between elements. The test statistic, given by
\begin{equation}
T_{CLX} = \max\limits_{1 \leq i < j \leq p} \frac{ \left( \mathcal{S}_{1, ij} - \mathcal{S}_{2, ij} \right)^2}{ \frac{ \omega_{1, ij}}{n} + \frac{\omega_{2, ij}}{m} },
\label{eqn:clxtest}
\end{equation}
where $\omega_{1, ij} = n^{-1} \sum\limits_{k=1}^n \left\{  (\xvec_{ki} - \overline{\xvec}_i) (\xvec_{kj} - \overline{\xvec}_j) - \mathcal{S}_{1, ij} \right\}^2$ and 
$\omega_{2, ij} = m^{-1} \sum\limits_{k=1}^m \left\{  (\yvec_{ki} - \overline{\yvec}_i) (\yvec_{kj} - \overline{\yvec}_j) - \mathcal{S}_{2, ij} \right\}^2$. Under $H_{0T}$, the limiting distribution of $T_{CLX}$ is shown be an extreme value distribution of type I. In comparison with the Frobenius norm based tests, $T_{CLX}$ is shown to be more powerful at detecting difference between the covariance matrices when the differences are {\it sparse}, i.e. they differ in very small number of elements.

\section{Projection based test}
\label{sec:proposedtest}

%Like two-sample mean vector testing, two-sample covariance testing is an important problem in multivariate analysis. The knowledge of equality of two covariance matrices is critical for many statistical procedures, including Fisher's linear discriminant analysis. The covariance structure also plays a central role in the analysis of data from many fields, including genomics, medical imaging,  and financial economics. However, the dimensions of data from these fields very often are much larger than sample sizes. Such a data set is termed as high-dimensional data.  Analysis of high-dimensional data has become one of the most important areas of current research in statistics. 

Conventional methods discussed for testing equality of covariance matrices usually fail in high-dimensional data settings because the sample covariance matrix does not converge to its population counterpart. Test statistics comparing covariance matrices are mainly based on matrix functions, such as eigenvalues, trace, Frobenius norm, etc., which also lose consistency in high dimensions. Thus performance of methods for comparison of covariance matrices {\it worsens} with increasing dimension. Test methods for covariance matrices in lower case enjoy many appealing properties. For example, $ U_{John}$ test is invariant and is also the locally most powerful. The high dimensional methods are shown perform well, but they fail to achieve the theoretical properties of $U_{John}$. The LRT in the two sample case is also robust and has good asymptotic properties when the dimension is smaller than the sample size. To preserve the properties of traditional multivariate methods, an attractive approach is to {\bf embed} the data and model into a lower dimension such that the hypothesis and inference are preserved. 
 
%{\color{red} Describe how lower-dimensional embeddings using a projection matrix preserve distances between points through Johnson-Lindenstrauss lemma}
When embedding data into lower-dimensional subspaces for parametric inference, the mapping should be such that the local topology of the data is preserved. Since the parameter of interest is the covariance matrix, which is a measure of spread, the mapping should preserve pairwise distances between observations. The existence of such a mapping is given by the Johnson-Lindenstrauss lemma \citep{Johnson1984}, which says that any linear mapping from the original space into the lower-dimensional space satisfies this condition. Hence we consider linear projection mappings from $\mathbb{R}^p$ into $\mathbb{R}^k$ for $k < p$ of the form $\xvec \mapsto \mathcal{R} \xvec$ where $\mathcal{R} \in \mathbb{R}^{k \times p}$ is the projection matrix. This paper's main motivation is to develop test methods for covariance matrices for high-dimensional data that enjoy the appealing properties of tests for covariance matrices for lower dimensional data. The most natural path to mimic the tests for covariance matrices for lower data, such as $ U_{John}$ test is to project high-dimensional data onto a space of dimension smaller than the sample size. 

When considering dimension reduction techniques, principal component analysis (PCA) is the most popular and commonly used.  While PCA is used very frequently for graphical representation and has good geometric properties, it is not ideal for projection-based hypothesis testing in high dimensions. For example, consider the two-sample test. When using PCA-based projection, covariance of the data projected onto the first $m$ principal component is given by the first $m$ eigenvalues. While the data is embedded in the lower dimension, the hypothesis is not preserved. Equality of the first $m$ eigenvalues does not guarantee that the two covariance matrices are equal. Extending to include all the $p$ eigenvalues will also not work since the sample covariance matrix is singular and yields only $n - 1$ non-zero eigenvalues. Other data-driven projection methods such as t-SNE \citep{tSNE} will also not work for similar reasons. To avoid these shortcomings, random projection (RP) of data is a popular method to alleviate the curse of dimensionality. 

A random projection matrix $R = (r_{ij}) \in \mathbb{R}^{m \times p}$ is a matrix with randomly generated elements, and is not generated from a matrix-valued distribution. The elements $r_{ij}$ are randomly and independently generated thereby resulting in a much lower computational cost. Structural constraints such as sparsity and orthogonality can be imposed later as desired. There are various methods to generate the elements of the random projection matrix - \citep{Achlioptas, Srivastava2014} generate {\it sparse} projection matrices by structuring the matrix to have a large proportion of zeros. Another approach is to impose structure by generating orthogonal matrices to preserve geometrical properties in the data. RP-based inference procedure is along the same lines as a union-intersection test, where the null hypothesis is equivalently written as the intersection of a family of hypotheses and the alternative is expressed as a union. The principle remains the same - we reject the null hypothesis if {\it at least one} random projection presents evidence in favor of rejection.

\subsection{Proposed test procedure}
First consider the one sample hypotheses. For $k < p$, let $\mathcal{R} \in \mathbb{R}^{k \times p}$ be a projection matrix and define $\xvecstar_i = \mathcal{R} \xvec_i, i = 1, \ldots n$ as the projected data. If the mean and variance of $\xvec$ are given by $\muvec$ and $\Sigma$ respectively, then we have $\muvec^* = \mathbb{E} (\xvecstar_i) = \mathcal{R} \muvec$ and $\Sigma^* = {\rm var}(\xvecstar_i) = \mathcal{R} \Sigma \mathcal{R}^{\top}$. Under the null hypothesis of identity, the variance of $\xvecstar$ becomes ${\rm var}(\xvecstar | H_{0I} ) = \mathcal{R} \Sigma \mathcal{R}^{\top} = \mathcal{R} \mathcal{R}^{\top}$. Similarly under the null hypothesis of sphericity, we have ${\rm var}(\xvecstar | H_{0S}) = \sigma^2 \mathcal{R} \mathcal{R}^{\top}$. If we choose the projection matrix $\mathcal{R}$ to be of full row rank and semi-orthogonal, i.e. $\mathcal{R} \mathcal{R}^{\top} = \mathcal{I}_k$, then the null hypotheses are preserved under the projection. Using $\xvecstar_1, \ldots, \xvecstar_n$ as the data, the hypotheses of interest will be
\begin{gather*}
H_{0I}^* : \Sigma^* = \mathcal{I}_k \hspace{5mm} \mbox{ vs. } \hspace{5mm} H_{1I} : \Sigma^* \neq \mathcal{I}_k, \\
H_{0S}^* : \Sigma^* = \sigma^2 \mathcal{I}_k \hspace{5mm} \mbox{ vs. } \hspace{5mm} H_{1S} : \Sigma^* \neq \sigma^2 \mathcal{I}_k. 
\end{gather*}

If the data $\xvec$ is assumed to follow a normal distribution, the projected observations $\xvecstar$ will also be normally distributed. Hence likelihood ratio tests can be used to test $H_{0I}^*$ and $H_{0S}^*$. Also, the functional based tests, $U_{John}$ and $V_{Nagao}$ in \eqref{eqn:johnnagao} can be used since the projection ensures $k < n$. Defining the sample covariance matrix $\mathcal{S}^* =  n^{-1} \sum\limits_{i = 1}^n \left(\xvecstar_i - \overline{\xvecstar} \right) \left( \xvecstar_i - \overline{\xvecstar} \right)^{\top}$, we have
\begin{equation}
U_{John}^* = \frac{1}{k} \tr \left\{ \frac{ \mathcal{S}^*}{\tr \mathcal{S}^*/k}  - \mathcal{I}_k \right\}^2, \hspace{1cm} V_{Nagao}^* = \frac{1}{k} \tr \left\{ \mathcal{S}^* - \mathcal{I}_k \right\}^2.
\label{eqn:johnnagao2}
\end{equation}
Asymptotically, these tests will have a chi-squared distribution with $\nu = k(k + 1)/2 - 1$ degrees of freedom. Hence the p-values are given by
\begin{equation}
\pi_U = \chi^2_{\nu} \left( U_{John}^* \right), \hspace{1cm} \pi_{V} =  \chi^2_{\nu} \left( V_{Nagao}^* \right),
\label{eqn:pvalues}
\end{equation}
which can be used to reject the null hypotheses.

The equivalence between $H_{0I}$ and $H_{0I}^*$ (similarly between $H_{0S}$ and $H_{0S}^*$) holds valid irrespective of the choice of the projection matrix $\mathcal{R}$.  Basing the inference on a single instance of $\mathcal{R}$ may lead to erroneous conclusions. For example, if we take $k = p/2$ and $\Sigma = \begin{bmatrix} \mathcal{I}_k & \mathbf{0} \\ \mathbf{0} & \Omega \end{bmatrix}$ for some symmetric positive definite matrix $\Omega$, then setting $\mathcal{R} = \begin{bmatrix} \mathcal{I}_k & \mathbf{0} \end{bmatrix}$ satisfies $H_{0I}^*$ but not $H_{0I}$. To avoid this issue, the cumulative decision based on multiple random projections needs to be considered. Combining the decisions of multiple random projections is a common practice when doing random projection based inference. In mean vector tests, \cite{Srivastava2014} used average $p$-values to combine the $M$ projections, while \cite{Wu2020} proposed using the maximum test statistic of the $M$ projections. We consider the average of $p$-values to make inference as the mean is more robust to {\it extreme} projections causing extreme $p$-values, although they have a very low probability of occurring. Let $\mathcal{R}_1, \ldots, \mathcal{R}_M$ be $M$ independent random projection matrices. Let $\pi_1, \ldots, \pi_M$ denote the respective p-values for the $m$ projections. We reject the null hypothesis if the average $p$-value is small,
\begin{equation*}
\overline{\pi} \leq q_{\alpha}, 
\end{equation*}
where $q_{\alpha}$ is the $\alpha$-level critical value of the sampling distribution of $\overline{\pi}$. 

The distribution of $\overline{\pi}$ is not known to compute $q_{\alpha}$. An asymptotic approximation for the distribution of $\overline{\pi}$ can be derived using the fact that the $p$-values are independent conditional on the observations.  However, such an approximation can introduce additional error into the test procedure. To avoid this error, critical values are computed by simulating the empirical distribution of $\overline{\pi}$ under the null hypothesis. Algorithm \ref{alg:algorithm1} outlines the test procedure for $H_{0S}$. For $H_{0I}$, the algorithm is similar with $U_{John}^*$ and $\pi_{U}$ replaced by $V_{Nagao}^*$ and $\pi_{V}$ respectively. 
\begin{algorithm}[ht]
	\SetAlgoLined
	\For{m = 1:M}{
		\For{k = 1:K}{
			Generate $\zvec_1, \ldots, \zvec_n$ under $H_{0S}$\;
		 	Generate $\mathcal{R}$ and compute $U_{John}^*$ and $\pi_{U, k}$ as in \eqref{eqn:pvalues}\;
	 	}
 	Compute $\overline{\pi}_m = {\rm mean}(\pi_{U, 1}, \ldots, \pi_{U, K})$\;
	}
	Return $\widehat{q}_{\alpha} = \overline{\pi}_{[U, M(1 - \alpha)]}$ as the empirical critical value
\caption{Generating the sampling distribution of the average of $p$-values to compute the empirical critical value for the one-sample tests}
\label{alg:algorithm1}
\end{algorithm}

Generating data under $H_{0I}$ is straightforward as the observations are generated from $\mathcal{N}_p \left( \mathbf{0}, \mathcal{I} \right)$. Under $H_{0S}$, the $\zvec$ are generated from $\mathcal{N}_p \left( \mathbf{0}, \sigma^2 \mathcal{I} \right)$ for some $\sigma \in \mathbb{R}$. As rejecting or accepting $H_{0S}$ is independent of the sphericity parameter, the choice of $\sigma$ should not affect the null distribution of $\overline{\pi}_{U}$. The following result establishes invariance of the distribution of $\overline{\pi}_{U}$ under $H_{0S}$. For practical implementation, the null distribution of $\overline{\pi}_U$ can therefore be constructed using Algorithm \ref{alg:algorithm1} by generating $\zvec_1, \ldots, \zvec_n$ from $\mathcal{N}(\bf{0}, \mathcal{I}_p)$. 
\begin{theorem}
	Let $\xvec_1, \ldots, \xvec_n$ be a random sample from $\mathcal{N}_p \left( \mathbf{0}, \sigma^2 \mathcal{I} \right)$. Let $U_{John}^*$ and $\pi_{U}$ be as defined in \eqref{eqn:pvalues}. Let $\mathcal{R}_1, \ldots, \mathcal{R}_M$ are independent random projection matrices of dimension $k \times p$ yielding p-values $\pi_1, \ldots, \pi_M$. If we define $\overline{\pi}_U$ as the mean of $\pi_1, \ldots, \pi_M$, then the distribution of $\overline{\pi}$ is independent of $\sigma$. 
	\label{thm:sphericity}
\end{theorem}
\noindent {\bf Proof}: See Appendix

\subsection{Two sample testing}
To test the equality of covariance matrices of two normal populations, the likelihood ratio test \eqref{eqn:lrt} or the Wald-type test \eqref{eqn:waldtest} can be used when $p < n + m$. For high-dimensional data, these tests can be applied by projecting the data into lower-dimensional subspace. For a random semi-orthogonal matrix $\mathcal{R} \in \mathbb{R}_{k \times p}$ of full row rank, let $\xvecstar_i = \mathcal{R} \xvec_i, i =1, \ldots, n$ and $\yvecstar_j = \mathcal{R} \yvec_j, j = 1, \ldots, m$ denote the projected observations from the two populations respectively. The hypothesis of equality of covariance matrices in \eqref{eqn:twosample} can be equivalently stated as $H_{0T} : \Sigma_1 - \Sigma_2  = 0$ versus $H_{1T} : \Sigma_1 - \Sigma_2 \neq 0$. In the projected subspace, the two-sample hypothesis will become
\begin{equation*}
H_{0T} : \mathcal{R} \left( \Sigma_1 - \Sigma_2 \right) \mathcal{R}^{\top}  = 0 \hspace{1cm} \mbox{ vs. } \hspace{1cm} H_{1T} : \mathcal{R} \left( \Sigma_1 - \Sigma_2 \right) \mathcal{R}^{\top}  \neq 0.
\end{equation*}
Let $\mathcal{S}_1^*, \mathcal{S}_2^*$ and $\mathcal{S}_{pl}^*$ denote the sample covariance matrices of the two groups and the pooled covariance matrix respectively. Then the projected Box-$M$ test statistic and the Wald-type test statistic will be
\begin{gather}
\begin{aligned}
\mathcal{M}^* &= \frac{ \left| \mathcal{S}_1^* \right|^{n - 1} \left| \mathcal{S}_2^*\right|^{m - 1}  }{ \left| \mathcal{S}_{pl}^* \right|^{n + m - 2} }, \\
T_{Wald}^* &= \frac{n + m}{2} \left[ \frac{n}{n + m} \tr \left( \mathcal{S}_1^* \mathcal{S}_{pl}^* \right)^2 + \frac{m}{n + m} \tr \left( \mathcal{S}_2^* \mathcal{S}_{pl}^{*^{-1}} \right)^2 - \frac{ n m}{(n + m)^2} \tr \left( \mathcal{S}_1^*  \mathcal{S}_{pl}^{*^{-1}} \mathcal{S}_2^* \mathcal{S}_{pl}^{*^{-1}} \right) \right].
\end{aligned}
\label{eqn:twosampletests}
\end{gather}
The p-values are calculated using the $\chi^2_{\eta}$ approximation with $\eta = k(k + 1)/2$. For $\mathcal{M}^*$, finite-sample correction terms as described in section \ref{sec:hypotheses} can be used to improve performance. 

As in the case of one-sample tests, the aggregate decision from multiple random projections should be used to accept or reject $H_{0T}$. For $m$ independent random projection matrices $\mathcal{R}_{\ell}, {\ell} = 1, \ldots, m$  with corresponding $p$-values $\pi_{\ell}$, let $\overline{\pi}$ denote the average $p$-value. To determine the $\alpha$-level critical value $q_{\alpha}$, the sampling distribution of $\overline{\pi}$ under $H_{0T}$ is required.  Under the null hypothesis, it is only known that the two covariance matrices are equal. Thus, the empirical sampling distribution can be generated using any $\Sigma_1 = \Sigma_2 = \Sigma$ for any symmetric positive definite matrix $\Sigma$. The following theorem provides invariance of the sampling distribution of $\overline{\pi}$ to the choice of parameters under $H_{0T}$.
\begin{theorem}
	\label{thm:twosample}
	Let $\xvec_i \sim \mathcal{N} \left(\muvec_1, \Sigma \right), i = 1, \ldots, n$ and $\yvec_j \sim \mathcal{N} \left( \muvec_2, \Sigma \right), j = 1, \ldots, m$ be two groups of independent observations. Let $\mathcal{M}^* $  be as defined in \eqref{eqn:twosampletests} and $\pi_{\ell}$ denote the $p$-value obtained when using the random projection $\mathcal{R}_{\ell}, \ell = 1, \ldots, m$. If $\overline{\pi}_{\mathcal{M}}$ denotes the average of the $m$ p-values, then the sampling distribution of $\overline{\pi}_{\mathcal{M}}$ is independent of $\muvec_1, \muvec_2$ and $\Sigma$.
\end{theorem}
\noindent {\bf Proof}: See Appendix

The above result indicates that random samples from standard normal distribution can be used to generate the empirical critical value. Implementation of the method is described in Algorithm \eqref{alg:algorithm2}. 
\begin{algorithm}[ht]
	\SetAlgoLined
	\For{m = 1:M}{
		\For{k = 1:K}{
			Generate $\zvec_{\xvec;1}, \ldots, \zvec_{\xvec;n}, \zvec_{\yvec;1}, \ldots, \zvec_{\yvec;m}  \sim \mathcal{N}_p (0, \mathcal{I}))$\;
			Generate $\mathcal{R}$ and project the data $\zvec^* = \mathcal{R} \zvec$\;
			Compute $\mathcal{S}_1^* = {\rm var}(\zvec_{\xvec;1}^*, \ldots, \zvec_{\xvec;n}^*)$ and $\mathcal{S}^*_2 = {\rm var}(\zvec_{\yvec;1}^*, \ldots, \zvec_{\yvec;m}^*)$\;
			Compute $\mathcal{M}^*$ and $\pi_{\mathcal{M}, k}$ as defined in \eqref{eqn:twosampletests}
		}
		Compute $\overline{\pi}_m = {\rm mean}(\pi_{\mathcal{M}, 1}, \ldots, \pi_{\mathcal{M}, K})$\;
	}
	Return $\widehat{q}_{\alpha} = \overline{\pi}_{[V, M(1 - \alpha)]}$ as the empirical critical value
	\caption{Generating the sampling distribution of the average of $p$-values to compute the empirical critical value for the two-sample tests}
	\label{alg:algorithm2}
\end{algorithm}

\section{Simulation study}
\label{sec:simulation}
To study the performance of the random projection based tests in comparison against the high-dimensional tests, we performed an extensive simulation study for both the one and two sample cases. Type I error and power are computed under different scenarios, for various values of sample sizes $n$ and $m$, dimension of the original sample space $p$ and projected spaces $M$, respectively. To study the effect of sample size and dimensions, we set $n \in \{20, 40, 50, 60\}$, $p \in \{ 100, 200, 500, 1000, 2000 \}$ and $M \in \{ 5, 10, 15\}$. Empirical size and power are computed at the nominal significance level of $\alpha = 0.05$.

\subsection{One sample results}
For the hypotheses of identity $H_{0I}$, we have the three test high dimensional test statistics - $V_{CZZ}, V_{LW}$ and $V_{SYK}$. We consider three random projection based tests - $LRT_{I}$, $V_{John}$ and $V_{LW}$. For all the studies, observations are randomly generated from a normal distribution with mean $\muvec$ and covariance matrix $\Sigma = (\sigma_{ij})_{1 \leq i, j \leq p}$. Elements of the mean vector were generated uniformly, $\mu_k \sim {\rm Unif}(-3, 3), i = 1, \ldots, p$. For computing type I error, the covariance matrix is set as identity matrix of dimension $p$. Power was computed under 4 scenarios (Power I -- Power IV) under the alternative, with the difference from identity matrix defined in two ways - a band matrix with non-zero diagonal elements and a diagonal matrix with elements different from 1. For Power I and II, we set $\sigma_{ij} = \rho^{|i - j|}$ for $|i - j| \leq B$ for some bandwidth $B$ and zero otherwise. For Power III and IV, we define $\Sigma$ as diagonal with $\sigma_{ii} = 1$ for $i \leq B$ and $\sigma_{ii} = 1 + \varepsilon$ for $B < i \leq p$. Table \ref{tab:onesamplesize} presents the type I error for $M = 5$ and $M = 15$. 

Among the high dimensional tests, only $V_{CZZ}$ preserves type I error at $5\%$ significance level. Both $T_{SYK}$ and $T_{LW}$ always reject the null hypothesis. When randomly projecting to $M = 5$ and $M = 15$ dimensions, all the three lower-dimensional tests control type I error rate, with the performance being slightly better for $M = 15$. Across all combinations of $n$ and $p$, the RP-based LRT and $V_{John}$ for both values of $M$ outperforms $V_{CZZ}$.  As $T_{SYK}$ and $T_{LW}$ fail to preserve type I error, only $T_{CZZ}$ and the lower dimensional tests are compared in the power studies for the four scenarios, results of which are presented in Table \ref{tab:onesamplepower}. In Power I and II, all the tests have comparable power for small dimensions ($p = 100, 200 ,500$). For fixed sample size, the power decreases with dimension. The power of the RP-based tests increase when the projected dimension $M$ is increased. For small sample size, $T_{LW}$ has higher power than $T_{LRT}$ and $T_{John}$ , with the LRT achieving higher power than $T_{LW}$ as $n$ increases to $50$. In Power III and IV, the random projection tests have greater power, with $T_{LW}$ outperforming all the tests. Overall, $T_{LW}$ with random projection seems to have the best performance across all the comparisons.
\begin{table}[!ht]
	\centering
	\begin{tabular}{*{11}{c}}
		\hline
		Sample size &	Dimension	&	\multicolumn{3}{c}{High dimension} & \multicolumn{3}{c}{$M = 5$} & \multicolumn{3}{c}{$M = 15$} \\
		\cmidrule(lr){3-5} \cmidrule(lr){6-8} \cmidrule(lr){9-11}
		& & $T_{CZZ}$	&	$T_{SYL}$	&	$T_{LW}$	&	LRT	&	$T_{LW}$	&	$V_{John}$	&	LRT	&	$T_{LW}$	&	$V_{John}$	\\
		\hline
		\multirow{5}{*}{N = 20}
		&	100	&	0.076	&	1	&	0.806	&	0.061	&	0.057	&	0.043	&	0.052	&	0.047	&	0.05	\\
		&	200	&	0.081	&	1	&	1	&	0.053	&	0.054	&	0.06	&	0.047	&	0.047	&	0.061	\\
		&	500	&	0.079	&	1	&	1	&	0.051	&	0.062	&	0.05	&	0.06	&	0.069	&	0.054	\\
		&	1000	&	0.062	&	1	&	1	&	0.054	&	0.05	&	0.055	&	0.061	&	0.056	&	0.052	\\
		&	2000	&	0.07	&	1	&	1	&	0.044	&	0.047	&	0.043	&	0.062	&	0.043	&	0.05	\\
		\hline
		\multirow{5}{*}{N = 40}
		&	100	&	0.068	&	1	&	0.358	&	0.055	&	0.049	&	0.059	&	0.049	&	0.042	&	0.063	\\
		&	200	&	0.061	&	1	&	0.785	&	0.05	&	0.041	&	0.049	&	0.06	&	0.052	&	0.053	\\
		&	500	&	0.058	&	1	&	1	&	0.051	&	0.048	&	0.056	&	0.059	&	0.067	&	0.062	\\
		&	1000	&	0.067	&	1	&	1	&	0.056	&	0.04	&	0.046	&	0.05	&	0.065	&	0.053	\\
		&	2000	&	0.054	&	1	&	1	&	0.072	&	0.054	&	0.048	&	0.056	&	0.049	&	0.052	\\
		\hline
		\multirow{5}{*}{N = 50}
	&	100	&	0.056	&	1	&	0.258	&	0.057	&	0.051	&	0.051	&	0.045	&	0.044	&	0.04	\\
	&	200	&	0.057	&	1	&	0.645	&	0.036	&	0.047	&	0.07	&	0.039	&	0.05	&	0.052	\\
	&	500	&	0.048	&	1	&	0.999	&	0.054	&	0.043	&	0.05	&	0.054	&	0.048	&	0.063	\\
	&	1000	&	0.061	&	1	&	1	&	0.065	&	0.054	&	0.05	&	0.074	&	0.064	&	0.038	\\
	&	2000	&	0.058	&	1	&	1	&	0.065	&	0.059	&	0.038	&	0.055	&	0.055	&	0.062	\\
		\hline
	\end{tabular}	
	\caption{Type I error of the three high dimensional tests ($T_{CZZ}$, $T_{SYL}$, $T_{LW}$) and the RP based tests for projected dimensions $M = 5$ and $M = 15$. The results are for various combinations of sample size and dimension, averaged over $1000$ replicates. }
	\label{tab:onesamplesize}	
\end{table}

\begin{sidewaystable}[!ht]
	\centering
	\begin{tabular}{cc |*{7}{c}|*{7}{c}}
		& & \multicolumn{7}{c}{Power I} & \multicolumn{7}{c}{Power II} \\
		\cmidrule(lr){3-9} \cmidrule(lr){10-16}
		& & & \multicolumn{3}{c}{$M = 5$}	& \multicolumn{3}{c}{$M = 15$} &  & \multicolumn{3}{c}{$M = 5$} & \multicolumn{3}{c}{$M = 15$} \\
		\cmidrule(lr){4-6} \cmidrule(lr){7-9} \cmidrule(lr){11-13} \cmidrule(lr){14-16}
		&	Dimension	&	$T_{CZZ}$	&	LRT	&	$T_{LW}$	&	$V_{John}$	&	LRT	&	$T_{LW}$	&	$V_{John}$	&	$T_{CZZ}$	&	LRT	&	$T_{LW}$	&	$V_{John}$	&	LRT	&	$T_{LW}$	&	$V_{John}$	\\
		\hline
		\multirow{5}{*}{$n = 20$}
		&	100	&	0.999	&	0.988	&	0.994	&	0.493	&	0.999	&	0.999	&	0.496	&	0.999	&	0.986	&	0.989	&	0.439	&	0.999	&	0.999	&	0.427	\\
		&	200	&	0.996	&	0.934	&	0.954	&	0.318	&	0.997	&	0.998	&	0.336	&	1	&	0.931	&	0.96	&	0.272	&	0.995	&	0.993	&	0.303	\\
		&	500	&	0.998	&	0.5	&	0.579	&	0.224	&	0.924	&	0.965	&	0.26	&	0.999	&	0.553	&	0.603	&	0.189	&	0.918	&	0.966	&	0.227	\\
		&	1000	&	0.999	&	0.195	&	0.246	&	0.155	&	0.573	&	0.79	&	0.177	&	1	&	0.26	&	0.281	&	0.183	&	0.556	&	0.732	&	0.161	\\
		&	2000	&	1	&	0.099	&	0.125	&	0.118	&	0.259	&	0.379	&	0.13	&	0.999	&	0.134	&	0.15	&	0.117	&	0.214	&	0.395	&	0.158	\\
			\hline
		\multirow{5}{*}{$n = 40$}
		&	100	&	1	&	1	&	1	&	0.987	&	1	&	1	&	0.974	&	1	&	1	&	1	&	0.988	&	1	&	1	&	0.963	\\
		&	200	&	1	&	1	&	1	&	0.924	&	1	&	1	&	0.907	&	1	&	1	&	1	&	0.92	&	1	&	1	&	0.897	\\
		&	500	&	1	&	0.98	&	0.98	&	0.626	&	1	&	1	&	0.701	&	1	&	0.986	&	0.988	&	0.65	&	1	&	1	&	0.697	\\
		&	1000	&	1	&	0.651	&	0.669	&	0.373	&	1	&	1	&	0.512	&	1	&	0.63	&	0.665	&	0.375	&	1	&	1	&	0.479	\\
		&	2000	&	1	&	0.304	&	0.347	&	0.228	&	0.783	&	0.907	&	0.354	&	1	&	0.242	&	0.245	&	0.217	&	0.792	&	0.893	&	0.342	\\
			\hline
		\multirow{5}{*}{$n = 50$}
		&	100	&	1	&	1	&	1	&	1	&	1	&	1	&	1	&	1	&	1	&	1	&	1	&	1	&	1	&	0.999	\\
		&	200	&	1	&	1	&	1	&	0.997	&	1	&	1	&	0.993	&	1	&	1	&	1	&	0.996	&	1	&	1	&	0.988	\\
		&	500	&	1	&	0.999	&	0.999	&	0.823	&	1	&	1	&	0.87	&	1	&	0.999	&	1	&	0.811	&	1	&	1	&	0.913	\\
		&	1000	&	1	&	0.76	&	0.772	&	0.512	&	1	&	1	&	0.657	&	1	&	0.83	&	0.847	&	0.514	&	1	&	1	&	0.669	\\
		&	2000	&	1	&	0.353	&	0.408	&	0.282	&	0.953	&	0.978	&	0.491	&	1	&	0.377	&	0.441	&	0.265	&	0.953	&	0.983	&	0.453	\\
		\hline
		& & \multicolumn{7}{c}{Power III} & \multicolumn{7}{c}{Power IV} \\
		\hline
		\multirow{5}{*}{$n = 20$}
		&	100	&	0.129	&	0.03	&	0.213	&	0.864	&	0.033	&	0.176	&	0.871	&	0.208	&	0.068	&	0.616	&	1	&	0.074	&	0.425	&	1	\\
		&	200	&	0.121	&	0.014	&	0.279	&	0.99	&	0.035	&	0.26	&	0.99	&	0.213	&	0.059	&	0.791	&	1	&	0.074	&	0.709	&	1	\\
		&	500	&	0.128	&	0.011	&	0.326	&	1	&	0.009	&	0.491	&	1	&	0.21	&	0.042	&	0.871	&	1	&	0.051	&	0.961	&	1	\\
		&	1000	&	0.136	&	0.008	&	0.322	&	1	&	0.02	&	0.578	&	1	&	0.181	&	0.036	&	0.918	&	1	&	0.039	&	0.991	&	1	\\
		&	2000	&	0.136	&	0.009	&	0.307	&	1	&	0.027	&	0.659	&	1	&	0.183	&	0.023	&	0.933	&	1	&	0.04	&	0.998	&	1	\\
		\hline
		\multirow{5}{*}{$n = 40$}		
		&	100	&	0.185	&	0.092	&	0.333	&	0.988	&	0.099	&	0.257	&	0.993	&	0.331	&	0.449	&	0.848	&	1	&	0.323	&	0.685	&	1	\\
		&	200	&	0.165	&	0.067	&	0.438	&	1	&	0.122	&	0.396	&	1	&	0.333	&	0.593	&	0.975	&	1	&	0.441	&	0.919	&	1	\\
		&	500	&	0.172	&	0.086	&	0.574	&	1	&	0.097	&	0.671	&	1	&	0.306	&	0.706	&	0.996	&	1	&	0.641	&	1	&	1	\\
		&	1000	&	0.17	&	0.061	&	0.546	&	1	&	0.068	&	0.73	&	1	&	0.285	&	0.742	&	1	&	1	&	0.672	&	1	&	1	\\
		&	2000	&	0.184	&	0.042	&	0.552	&	1	&	0.057	&	0.803	&	1	&	0.312	&	0.707	&	1	&	1	&	0.773	&	1	&	1	\\
		hline
		\multirow{5}{*}{$n = 50$}		
		&	100	&	0.208	&	0.182	&	0.452	&	0.995	&	0.159	&	0.328	&	0.996	&	0.366	&	0.708	&	0.94	&	1	&	0.457	&	0.802	&	1	\\
		&	200	&	0.203	&	0.136	&	0.597	&	1	&	0.169	&	0.544	&	1	&	0.354	&	0.846	&	0.993	&	1	&	0.681	&	0.967	&	1	\\
		&	500	&	0.179	&	0.107	&	0.608	&	1	&	0.157	&	0.741	&	1	&	0.353	&	0.925	&	1	&	1	&	0.888	&	1	&	1	\\
		&	1000	&	0.189	&	0.079	&	0.571	&	1	&	0.182	&	0.88	&	1	&	0.37	&	0.957	&	1	&	1	&	0.947	&	1	&	1	\\
		&	2000	&	0.18	&	0.075	&	0.655	&	1	&	0.111	&	0.87	&	1	&	0.383	&	0.955	&	1	&	1	&	0.954	&	1	&	1	\\
		hline
		\end{tabular}
	\caption{Power of the three high dimensional tests ($T_{CZZ}$, $T_{SYL}$, $T_{LW}$) and the RP based tests for projected dimensions $M = 5$ and $M = 15$. The results are for various combinations of sample size and dimension under four different scenarios averaged over $1000$ replicates.}
	\label{tab:onesamplepower}
\end{sidewaystable}

\subsection{Two sample results}
For the two sample test in equation \eqref{eqn:twosample}, we have four high dimensional tests -  $T_{Sch}, T_{SYK}, T_{LC}$ and $T_{CLX}$ and two standard dimension tests - Box's $M$ and Wald's test. All the random samples are generated from $p$-dimensional normal distributions with means $\muvec_1 = \muvec_2 = \mathbf{0}$ and covariance matrices $\Sigma_1$ and $\Sigma_2$ respectively . For type I error, we set both $\Sigma_1$ and $\Sigma_2$ to be the identity matrix. The results are presented in Table \ref{tab:twosamplesize}. We considered a total of 8 settings  (Power I -- Power VIII) to compare the power of the high dimensional tests and the RP-based tests. We considered two models for differentiating the covariance matrices - unequal values along the diagonal and band matrices. For Power I--IV, we set $\Sigma = \mathcal{I}_p$ and $\Sigma_2 = {\rm diag}(\sigma_{21}, \ldots, \sigma_{2p})$, where $\sigma_{2k} = 1$ for $k \leq [Bp]$ and $\sigma_{2k} \sim \Gamma(4,2)$ for $k = [Bp]+1, \ldots, p$. The bandwidth $B$ is varied over the 4 scenarios. For Power V--VIII, we set $\Sigma = {\rm diag}(\sigma_{11}, \ldots, \sigma_{1p})$ with $\sigma_{1k} \sim {\rm Unif}(1, 3)$ and $\Sigma_2 = \Sigma_1^{1/2} \Omega \Sigma_1^{1/2}$, where $\Omega$ is set as a band matrix with $\Omega_{ij} = \rho^{|i - j|}$ for $|i - j| \leq Bp$ and 0 otherwise. The parameter $B$ determines the width of the band matrix $\Omega$. 

Results for the type I error comparison are presented in Table \ref{tab:twosamplesize}. At the nominal $5 \%$ significance level, none of the high dimensional tests preserve type I error for the chosen combinations of $p$ and $n$. Amongst the RP-tests, both the Box's $M$-test and Wald test after random projections consistently preserves type I error rate for all values of $M$. It is interesting to note that the Wu-Li test, which is also based on random projections onto one dimension, fails to control type I error. This indicates that RP-based work well so long as the projected dimension is not very low. Tables \ref{tab:twosamplepower} and \ref{tab:twosamplepower2} present the power of the Box's $M$-test and Wald test respectively for the eight power scenarios. We did not include the high dimensional methods as they failed to control type I error. For all the eight scenarios, the RP-based tests seems to be achieve reasonable power, with the power decreasing with increase in $M$ and $p$. For any given values of $n, p, M$ and band size $B$, alternatives based on the band matrix achieve more power than differences along the diagonal. 
\begin{table}[!ht]
	\centering
	\begin{tabular}{*{13}{c}}
		& & \multicolumn{5}{c}{High dimension} & \multicolumn{3}{c}{Box test}	&	\multicolumn{3}{c}{Wald test} \\			
		\cmidrule(lr){3-7} \cmidrule(lr){8-10} \cmidrule(lr){11-13}
		&	Dimension	&	$T_{CLX}$	&	$T_{SYK}$	&	$T_{LC}$	&	$T_{Sch}$	&	WuLi	&	M = 5	&	10	&	15	&	M = 5	&	10	&	15	\\
		\hline
		\multirow{5}{*}{$n = 20$}																							
		&	100	&	0.85	&	1	&	0.229	&	0.454	&	0.973	&	0.054	&	0.052	&	0.053	&	0.052	&	0.047	&	0.045	\\
		&	200	&	0.838	&	1	&	0.185	&	0.433	&	0.97	&	0.053	&	0.039	&	0.045	&	0.049	&	0.035	&	0.059	\\
		&	500	&	0.758	&	1	&	0.219	&	0.466	&	0.971	&	0.049	&	0.037	&	0.042	&	0.043	&	0.055	&	0.068	\\
		&	1000	&	0.711	&	1	&	0.224	&	0.482	&	0.976	&	0.049	&	0.04	&	0.046	&	0.056	&	0.048	&	0.062	\\
		&	2000	&	0.616	&	1	&	0.214	&	0.504	&	0.98	&	0.035	&	0.056	&	0.038	&	0.035	&	0.047	&	0.052	\\
		
		\hline
		\multirow{5}{*}{$n = 40$}		
		&	100	&	0.926	&	1	&	0.236	&	0.434	&	0.969	&	0.059	&	0.069	&	0.062	&	0.027	&	0.036	&	0.035	\\
		&	200	&	0.934	&	1	&	0.2	&	0.439	&	0.976	&	0.053	&	0.054	&	0.054	&	0.061	&	0.059	&	0.068	\\
		&	500	&	0.929	&	1	&	0.224	&	0.468	&	0.97	&	0.053	&	0.04	&	0.058	&	0.051	&	0.047	&	0.042	\\
		&	1000	&	0.921	&	1	&	0.207	&	0.474	&	0.976	&	0.052	&	0.038	&	0.043	&	0.052	&	0.045	&	0.052	\\
		&	2000	&	0.895	&	1	&	0.224	&	0.504	&	0.97	&	0.059	&	0.067	&	0.045	&	0.062	&	0.065	&	0.044	\\
		
		\hline
		\multirow{5}{*}{$n = 50$}		
		&	100	&	0.939	&	1	&	0.243	&	0.45	&	0.968	&	0.047	&	0.047	&	0.047	&	0.056	&	0.053	&	0.053	\\
		&	200	&	0.944	&	1	&	0.215	&	0.445	&	0.973	&	0.042	&	0.035	&	0.046	&	0.056	&	0.034	&	0.045	\\
		&	500	&	0.929	&	1	&	0.225	&	0.501	&	0.968	&	0.041	&	0.063	&	0.066	&	0.052	&	0.043	&	0.039	\\
		&	1000	&	0.936	&	1	&	0.196	&	0.507	&	0.974	&	0.041	&	0.04	&	0.044	&	0.059	&	0.043	&	0.029	\\
		&	2000	&	0.934	&	1	&	0.238	&	0.485	&	0.976	&	0.052	&	0.036	&	0.033	&	0.033	&	0.038	&	0.04	\\
		
		\hline
		\multirow{5}{*}{$n = 60$}		
		&	100	&	0.946	&	1	&	0.215	&	0.434	&	0.968	&	0.044	&	0.054	&	0.052	&	0.036	&	0.044	&	0.044	\\
		&	200	&	0.953	&	1	&	0.234	&	0.433	&	0.964	&	0.046	&	0.037	&	0.049	&	0.05	&	0.049	&	0.06	\\
		&	500	&	0.947	&	1	&	0.224	&	0.477	&	0.973	&	0.054	&	0.047	&	0.036	&	0.047	&	0.047	&	0.032	\\
		&	1000	&	0.933	&	1	&	0.21	&	0.481	&	0.968	&	0.061	&	0.05	&	0.049	&	0.061	&	0.06	&	0.065	\\
		&	2000	&	0.933	&	1	&	0.209	&	0.497	&	0.969	&	0.053	&	0.034	&	0.077	&	0.066	&	0.041	&	0.059	\\
		
	\end{tabular}
	\caption{Type I error of the four high dimensional tests ($T_{CLX}$, $T_{SYK}$, $T_{CW}$, $T_{Sch}$) and the RP based tests for projected dimensions $M = 5, 10$ and $M = 15$. The results are for various combinations of sample size and dimension, averaged over $1000$ replicates. }
	\label{tab:twosamplesize}
\end{table}
		
\begin{table}[!ht]
	\centering
	\begin{tabular}{*{14}{c}}
			& & \multicolumn{3}{c}{Power I}	&	\multicolumn{3}{c}{Power II}  &  \multicolumn{3}{c}{Power III}	&	\multicolumn{3}{c}{Power IV} \\
			\cmidrule(lr){3-5} \cmidrule(lr){6-8} \cmidrule(lr){9-11} \cmidrule(lr){12-14}
		&	$M \rightarrow$	&	5	&	10	&	15	&	5	&	10	&	15	&	5	&	10	&	15	&	5	&	10	&	15	\\
		&	Dimension	&		&		&		&		&		&		&		&		&		&		&		&		\\
		\hline
		\multirow{5}{*}{$n = 20$}		
		&	100	&	0.644	&	0.523	&	0.43	&	0.583	&	0.428	&	0.301	&	0.918	&	0.738	&	0.511	&	0.999	&	0.975	&	0.835	\\
		&	200	&	0.427	&	0.314	&	0.259	&	0.808	&	0.629	&	0.478	&	0.949	&	0.816	&	0.565	&	1	&	0.982	&	0.919	\\
		&	500	&	0.425	&	0.313	&	0.181	&	0.945	&	0.848	&	0.554	&	0.999	&	0.985	&	0.876	&	1	&	1	&	0.997	\\
		&	1000	&	0.306	&	0.236	&	0.142	&	0.954	&	0.841	&	0.552	&	1	&	1	&	0.963	&	1	&	1	&	1	\\
		&	2000	&	0.207	&	0.155	&	0.125	&	0.952	&	0.791	&	0.517	&	1	&	0.996	&	0.9	&	1	&	1	&	0.997	\\
		\hline
		\multirow{5}{*}{$n = 40$}
		&	100	&	0.488	&	0.402	&	0.303	&	1	&	0.996	&	0.981	&	1	&	0.995	&	0.962	&	1	&	1	&	1	\\
		&	200	&	0.36	&	0.336	&	0.309	&	1	&	1	&	0.998	&	1	&	1	&	1	&	1	&	1	&	1	\\
		&	500	&	0.59	&	0.56	&	0.456	&	1	&	1	&	1	&	1	&	1	&	1	&	1	&	1	&	1	\\
		&	1000	&	0.697	&	0.654	&	0.59	&	1	&	1	&	1	&	1	&	1	&	1	&	1	&	1	&	1	\\
		&	2000	&	0.653	&	0.645	&	0.539	&	1	&	1	&	1	&	1	&	1	&	1	&	1	&	1	&	1	\\
		\hline
		\multirow{5}{*}{$n = 40$}
		&	100	&	0.935	&	0.827	&	0.738	&	0.901	&	0.755	&	0.617	&	1	&	1	&	0.998	&	1	&	1	&	0.996	\\
		&	200	&	0.902	&	0.838	&	0.795	&	1	&	0.999	&	0.997	&	1	&	1	&	1	&	1	&	1	&	1	\\
		&	500	&	0.777	&	0.826	&	0.813	&	1	&	1	&	1	&	1	&	1	&	1	&	1	&	1	&	1	\\
		&	1000	&	0.946	&	0.972	&	0.947	&	1	&	1	&	1	&	1	&	1	&	1	&	1	&	1	&	1	\\
		&	2000	&	0.807	&	0.795	&	0.776	&	1	&	1	&	1	&	1	&	1	&	1	&	1	&	1	&	1	\\
		\hline
		\multirow{5}{*}{$n = 60$}
		&	100	&	0.95	&	0.875	&	0.779	&	1	&	1	&	1	&	1	&	1	&	1	&	1	&	1	&	1	\\
		&	200	&	0.77	&	0.718	&	0.653	&	1	&	1	&	1	&	1	&	1	&	1	&	1	&	1	&	1	\\
		&	500	&	0.962	&	0.952	&	0.957	&	1	&	1	&	1	&	1	&	1	&	1	&	1	&	1	&	1	\\
		&	1000	&	0.887	&	0.93	&	0.89	&	1	&	1	&	1	&	1	&	1	&	1	&	1	&	1	&	1	\\
		&	2000	&	0.96	&	0.952	&	0.92	&	1	&	1	&	1	&	1	&	1	&	1	&	1	&	1	&	1	\\
		\hline
		\hline
		& & \multicolumn{3}{c}{Power V}	&	\multicolumn{3}{c}{Power VI}  &  \multicolumn{3}{c}{Power VII}	&	\multicolumn{3}{c}{Power VIII} \\
		\cmidrule(lr){3-5} \cmidrule(lr){6-8} \cmidrule(lr){9-11} \cmidrule(lr){12-14}
		&	$M	\rightarrow$ &	5	&	10	&	15	&	5	&	10	&	15	&	5	&	10	&	15	&	5	&	10	&	15	\\		
		&	Dimension	&		&		&		&		&		&		&		&		&		&		&		&		\\								
		\hline
		\multirow{5}{*}{$n = 20$}
		&	100	&	1	&	1	&	1	&	1	&	1	&	1	&	1	&	1	&	1	&	1	&	1	&	1	\\
		&	200	&	0.998	&	1	&	0.999	&	0.995	&	1	&	1	&	0.999	&	1	&	1	&	0.999	&	1	&	1	\\
		&	500	&	0.763	&	0.963	&	0.971	&	0.757	&	0.941	&	0.968	&	0.718	&	0.96	&	0.98	&	0.758	&	0.968	&	0.975	\\
		&	1000	&	0.359	&	0.639	&	0.635	&	0.354	&	0.642	&	0.679	&	0.33	&	0.62	&	0.671	&	0.401	&	0.682	&	0.671	\\
		&	2000	&	0.138	&	0.231	&	0.314	&	0.153	&	0.241	&	0.294	&	0.181	&	0.272	&	0.318	&	0.162	&	0.252	&	0.33	\\
		\hline
		\multirow{5}{*}{$n = 40$}
		&	100	&	1	&	1	&	1	&	1	&	1	&	1	&	1	&	1	&	1	&	1	&	1	&	1	\\
		&	200	&	1	&	1	&	1	&	1	&	1	&	1	&	1	&	1	&	1	&	1	&	1	&	1	\\
		&	500	&	1	&	1	&	1	&	1	&	1	&	1	&	1	&	1	&	1	&	1	&	1	&	1	\\
		&	1000	&	0.874	&	1	&	1	&	0.865	&	1	&	1	&	0.895	&	1	&	1	&	0.888	&	1	&	1	\\
		&	2000	&	0.368	&	0.859	&	0.959	&	0.342	&	0.811	&	0.951	&	0.424	&	0.779	&	0.939	&	0.432	&	0.822	&	0.942	\\
		\hline
		\multirow{5}{*}{$n = 50$}
		&	100	&	1	&	1	&	1	&	1	&	1	&	1	&	1	&	1	&	1	&	1	&	1	&	1	\\
		&	200	&	1	&	1	&	1	&	1	&	1	&	1	&	1	&	1	&	1	&	1	&	1	&	1	\\
		&	500	&	1	&	1	&	1	&	1	&	1	&	1	&	1	&	1	&	1	&	1	&	1	&	1	\\
		&	1000	&	0.968	&	1	&	1	&	0.947	&	1	&	1	&	0.963	&	1	&	1	&	0.975	&	1	&	1	\\
		&	2000	&	0.538	&	0.956	&	0.996	&	0.536	&	0.949	&	0.999	&	0.597	&	0.957	&	0.997	&	0.566	&	0.955	&	0.998	\\
		\hline
		\multirow{5}{*}{$n = 60$}
		&	100	&	1	&	1	&	1	&	1	&	1	&	1	&	1	&	1	&	1	&	1	&	1	&	1	\\
		&	200	&	1	&	1	&	1	&	1	&	1	&	1	&	1	&	1	&	1	&	1	&	1	&	1	\\
		&	500	&	1	&	1	&	1	&	1	&	1	&	1	&	1	&	1	&	1	&	1	&	1	&	1	\\
		&	1000	&	0.994	&	1	&	1	&	0.998	&	1	&	1	&	0.992	&	1	&	1	&	0.996	&	1	&	1	\\
		&	2000	&	0.734	&	0.992	&	1	&	0.694	&	0.996	&	1	&	0.761	&	0.987	&	1	&	0.692	&	0.99	&	1	\\
		\hline
	\end{tabular}
	\caption{Power of the RP based Box-$M$ test for projected dimensions $M = 5, 10$ and $M = 15$. The results are for various combinations of sample size and dimension under different scenarios averaged over $1000$ replicates. }
	\label{tab:twosamplepower}
\end{table}

\begin{table}[!ht]
	\centering
	\begin{tabular}{*{14}{c}}
		& & \multicolumn{3}{c}{Power I}	&	\multicolumn{3}{c}{Power II}  &  \multicolumn{3}{c}{Power III}	&	\multicolumn{3}{c}{Power IV} \\
		\cmidrule(lr){3-5} \cmidrule(lr){6-8} \cmidrule(lr){9-11} \cmidrule(lr){12-14}
		&	$M \rightarrow$	&	5	&	10	&	15	&	5	&	10	&	15	&	5	&	10	&	15	&	5	&	10	&	15	\\
		&	Dimension	&		&		&		&		&		&		&		&		&		&		&		&		\\
		\hline
		\multirow{5}{*}{$n = 20$}		
		&	100	&	0.252	&	0.18	&	0.153	&	0.931	&	0.751	&	0.569	&	0.89	&	0.709	&	0.515	&	0.998	&	0.947	&	0.791	\\
		&	200	&	0.297	&	0.245	&	0.216	&	0.733	&	0.482	&	0.374	&	0.998	&	0.933	&	0.768	&	1	&	0.998	&	0.972	\\
		&	500	&	0.374	&	0.24	&	0.219	&	0.723	&	0.55	&	0.37	&	1	&	0.983	&	0.921	&	1	&	1	&	0.995	\\
		&	1000	&	0.244	&	0.177	&	0.184	&	0.929	&	0.725	&	0.486	&	1	&	0.999	&	0.969	&	1	&	1	&	0.999	\\
		&	2000	&	0.218	&	0.173	&	0.173	&	0.825	&	0.595	&	0.366	&	1	&	0.99	&	0.917	&	1	&	1	&	0.994	\\		
		\hline
		\multirow{5}{*}{$n = 40$}
		&	100	&	0.802	&	0.697	&	0.567	&	0.994	&	0.959	&	0.889	&	1	&	1	&	0.998	&	1	&	1	&	1	\\
		&	200	&	0.721	&	0.681	&	0.564	&	1	&	0.991	&	0.965	&	1	&	1	&	1	&	1	&	1	&	1	\\
		&	500	&	0.621	&	0.565	&	0.463	&	1	&	0.999	&	0.996	&	1	&	1	&	1	&	1	&	1	&	1	\\
		&	1000	&	0.404	&	0.408	&	0.292	&	1	&	1	&	1	&	1	&	1	&	1	&	1	&	1	&	1	\\
		&	2000	&	0.695	&	0.627	&	0.524	&	1	&	1	&	1	&	1	&	1	&	1	&	1	&	1	&	1	\\		
		\hline
		\multirow{5}{*}{$n = 40$}
		&	100	&	0.907	&	0.856	&	0.745	&	0.998	&	0.994	&	0.965	&	0.999	&	0.997	&	0.983	&	1	&	1	&	0.999	\\
		&	200	&	0.706	&	0.629	&	0.572	&	1	&	1	&	1	&	1	&	1	&	1	&	1	&	1	&	1	\\
		&	500	&	0.813	&	0.791	&	0.757	&	1	&	1	&	1	&	1	&	1	&	1	&	1	&	1	&	1	\\
		&	1000	&	0.702	&	0.721	&	0.575	&	1	&	1	&	1	&	1	&	1	&	1	&	1	&	1	&	1	\\
		&	2000	&	0.761	&	0.701	&	0.646	&	1	&	1	&	1	&	1	&	1	&	1	&	1	&	1	&	1	\\		
		\hline
		\multirow{5}{*}{$n = 60$}
		&	100	&	0.901	&	0.812	&	0.71	&	0.903	&	0.825	&	0.725	&	1	&	1	&	1	&	1	&	1	&	1	\\
		&	200	&	0.995	&	0.993	&	0.974	&	1	&	1	&	1	&	1	&	1	&	1	&	1	&	1	&	1	\\
		&	500	&	0.979	&	0.983	&	0.964	&	1	&	1	&	1	&	1	&	1	&	1	&	1	&	1	&	1	\\
		&	1000	&	0.953	&	0.953	&	0.945	&	1	&	1	&	1	&	1	&	1	&	1	&	1	&	1	&	1	\\
		&	2000	&	0.902	&	0.879	&	0.857	&	1	&	1	&	1	&	1	&	1	&	1	&	1	&	1	&	1	\\		
		\hline
		\hline
		& & \multicolumn{3}{c}{Power V}	&	\multicolumn{3}{c}{Power VI}  &  \multicolumn{3}{c}{Power VII}	&	\multicolumn{3}{c}{Power VIII} \\
		\cmidrule(lr){3-5} \cmidrule(lr){6-8} \cmidrule(lr){9-11} \cmidrule(lr){12-14}
		&	$M	\rightarrow$ &	5	&	10	&	15	&	5	&	10	&	15	&	5	&	10	&	15	&	5	&	10	&	15	\\		
		&	Dimension	&		&		&		&		&		&		&		&		&		&		&		&		\\								
		\hline
		\multirow{5}{*}{$n = 20$}
		&	100	&	1	&	1	&	1	&	1	&	1	&	1	&	0.999	&	1	&	1	&	1	&	1	&	1	\\
		&	200	&	0.998	&	1	&	1	&	0.998	&	1	&	1	&	0.997	&	1	&	1	&	0.996	&	1	&	1	\\
		&	500	&	0.808	&	0.972	&	0.988	&	0.737	&	0.968	&	0.992	&	0.766	&	0.978	&	0.993	&	0.749	&	0.967	&	0.994	\\
		&	1000	&	0.343	&	0.676	&	0.787	&	0.403	&	0.638	&	0.789	&	0.35	&	0.644	&	0.795	&	0.372	&	0.654	&	0.778	\\
		&	2000	&	0.154	&	0.296	&	0.401	&	0.125	&	0.278	&	0.352	&	0.162	&	0.261	&	0.373	&	0.168	&	0.289	&	0.392	\\		
		\hline
		\multirow{5}{*}{$n = 40$}
		&	100	&	1	&	1	&	1	&	1	&	1	&	1	&	1	&	1	&	1	&	1	&	1	&	1	\\
		&	200	&	1	&	1	&	1	&	1	&	1	&	1	&	1	&	1	&	1	&	1	&	1	&	1	\\
		&	500	&	0.999	&	1	&	1	&	1	&	1	&	1	&	1	&	1	&	1	&	1	&	1	&	1	\\
		&	1000	&	0.858	&	1	&	1	&	0.892	&	0.998	&	1	&	0.862	&	1	&	1	&	0.885	&	0.999	&	1	\\
		&	2000	&	0.381	&	0.839	&	0.969	&	0.415	&	0.821	&	0.963	&	0.351	&	0.82	&	0.961	&	0.415	&	0.829	&	0.986	\\		
		\hline
		\multirow{5}{*}{$n = 50$}
		&	100	&	1	&	1	&	1	&	1	&	1	&	1	&	1	&	1	&	1	&	1	&	1	&	1	\\
		&	200	&	1	&	1	&	1	&	1	&	1	&	1	&	1	&	1	&	1	&	1	&	1	&	1	\\
		&	500	&	1	&	1	&	1	&	1	&	1	&	1	&	1	&	1	&	1	&	1	&	1	&	1	\\
		&	1000	&	0.97	&	1	&	1	&	0.962	&	1	&	1	&	0.967	&	1	&	1	&	0.966	&	1	&	1	\\
		&	2000	&	0.595	&	0.944	&	0.997	&	0.489	&	0.931	&	0.999	&	0.59	&	0.968	&	0.998	&	0.566	&	0.961	&	0.997	\\		
		\hline
		\multirow{5}{*}{$n = 60$}
		&	100	&	1	&	1	&	1	&	1	&	1	&	1	&	1	&	1	&	1	&	1	&	1	&	1	\\
		&	200	&	1	&	1	&	1	&	1	&	1	&	1	&	1	&	1	&	1	&	1	&	1	&	1	\\
		&	500	&	1	&	1	&	1	&	1	&	1	&	1	&	1	&	1	&	1	&	1	&	1	&	1	\\
		&	1000	&	0.996	&	1	&	1	&	0.996	&	1	&	1	&	1	&	1	&	1	&	0.998	&	1	&	1	\\
		&	2000	&	0.754	&	0.988	&	1	&	0.677	&	0.993	&	1	&	0.692	&	0.993	&	1	&	0.748	&	0.995	&	1	\\
		\hline
	\end{tabular}
	\caption{Power of the RP based Wald test for projected dimensions $M = 5, 10$ and $M = 15$. The results are for various combinations of sample size and dimension under different scenarios averaged over $1000$ replicates.}
	\label{tab:twosamplepower2}
\end{table}

\section{Data analysis}
\label{sec:realdata}
To study how the RP-based tests and the high dimensional test statistics perform when applied to real data, we considered two data sets. The first data set is a gene expression data from 62 colon tissues -  $n = 22$ normal and $m = 40$ tumor samples \citep{Alon1999}. Gene expression intensities of $p = 2000$ genes with highest minimal intensity were reported \footnote{\href{http://genomics-pubs.princeton.edu/oncology/affydata/index.html}{http://genomics-pubs.princeton.edu/oncology/affydata/index.html}}. We refer to this data set as \texttt{colon} henceforth. For the second illustration, we have gathered data on breast cancer subjects from the cancer genome atlas (TCGA) \footnote{\href{https://portal.gdc.cancer.gov/}{https://portal.gdc.cancer.gov/}}. Gene expression data from the RNA-Seq protocol are downloaded for patients from Stages IA, IIB and IIIC, resulting in samples of sizes 91, 291 and 70 respectively. The top $p = 2000$ genes with highest minimal intensity are kept in the final data set, which will be called \texttt{breast} henceforth.

\subsection{\texttt{colon} data} 
For the \texttt{colon} data, we did two analyses to compare the type I error rate and power of the test statistics in detecting differences in covariance matrices. First, the $n = 40$ tumor samples were randomly divided into two equal groups and tested for equality of covariance matrices. Since the sub-samples are from the same population, we expect the tests to not detect a significant difference between the covariance matrices of the two groups. We repeated this process $N = 1000$ times and the average number of false rejections is calculated. Second, we compared the normal and tumor samples.  It is widely accepted that in addition to the signals, co-expression networks are also not to vary with disease status. Hence we expect to detect a significance difference between the two covariance matrices.  Results are presented in Table \ref{tab:colon}. The type I error calculations indicate that the random projection tests do not falsely reject the null hypothesis, whereas $T_{SYK}$ and Wu-Li tests have a very high type I error. Only $T_{Sch}$ and the RP-based tests have very low type I error rate and correctly identify the difference between tumor and normal samples. While $T_{CLX}$ also correctly fails to reject $H_0$ under the null and $T_{LC}$ also controls type I error reasonably, their conclusion on comparison between the two groups (fail to reject) does not match the conclusion of the RP-based tests. All the methods have identified a difference in covariance structures between normal and tumor samples. 
\begin{table}[ht]
	\centering
	\begin{tabular}{C{3cm}|C{2cm}|C{3cm}}
		\hline
		Test & Type I error & Colon vs. Tumor p-value/Decision \\
		\hline
		$T_{SYK}$ & 0.543 & 0.0006 \\
		$T_{Sch}$ & 0.035 & 0 \\
		WuLi & 0.974 & Reject $H_0$ \\
		$T_{CLX}$ & 0.001 & Do not reject $H_0$ \\
		$T_{LC}$ & 0.041 & Do not Reject $H_0$ \\
		Box test - $M = 5$ & 0 & Reject $H_0$ \\
		Box test - $M = 10$ & 0 & Reject $H_0$ \\
	    Wald test - $M = 5$ & 0 & Reject $H_0$ \\
	    Wald test - $M = 10$ & 0 & Reject $H_0$  \\
	    \hline
	\end{tabular}
\caption{Results for type I error comparing sub-samples within the tumor samples and power for comparison between tumor and colon samples from the \texttt{colon} data set. The results are based on $1000$ bootstrap samples. }
\label{tab:colon}
\end{table}

\subsection{\texttt{breast} data}
In the \texttt{breast} data, the samples are divided into three groups based on the cancer stage. Similar to the \texttt{colon} data, we compared both type I error and power of the tests. First, we compared the type I error within each stage. Two samples of size 40 each are drawn to represent the two groups of observations. Since the observations correspond to the same stage, we expect the tests to not reject the null hypothesis. Proportion of rejections in $N = 1000$ repetitions will indicate the type I error within each cancer stage. Second, we compared the power of detecting difference between the stages. Using samples from different stages, power of the tests are similarly calculated. Results for both type I error and power are presented in Table \ref{tab:breast}. All the high dimensional methods have inflated type I error rates whereas the RP-based Box $M$-test and Wald test have very low false positives for stages IA and IIIC. It is interesting to note that for Stage IIB, all the test procedures have inflated type I error including the RP-based tests. This is a strong indication that there is potentially high heterogeneity within the samples resulting in the tests being rejected. The RP-based tests achieve very high power when comparing between the cancer stages. 
\begin{table}[!ht]
	\centering
	\begin{tabular}{c|*{3}{c}|C{2cm}C{2cm}C{2cm}}
		\hline
		\multirow{2}{*}{Test} & \multicolumn{3}{c|}{Type I error} & \multicolumn{3}{c}{Power} \\
		        &  Stage IA & Stage IIB & Stage IIIC & Stage IA vs. Stage IIB & Stage IA vs. Stage IIIC & Stage IIB vs. Stage IIIC \\
		        \hline
		        $T_{SYK}$ & 0.999 & 1 & 1 & 1 & 1 & 1 \\
		        $T_{Sch}$ & 0.285 & 0.541 & 0 & 0.986 & 0.986 & 0.842 \\
		        WuLi & 0.969 & 0.971 & 0.974 & 0.964 & 0.976 & 0.964 \\ 
		        $T_{CLX}$ & 0  & 0.004 & 0 & 0.02 & 0.066 & 0.034 \\
		        $T_{LC}$ & 0.002 & 0.156 & 0 & 0.492 & 0.696 & 0.399 \\
		        Box test - $M  = 5$ & 0.056 & 0.332 & 0 & 0.97 & 0.988 & 0.914 \\
		        Box test - $M = 10$ & 0.019 & 0.468 & 0 & 1 & 1 & 0.992 \\
		        Wald test - $M = 5$ & 0.046 & 0.289 & 0 & 0.96 & 0.98 & 0.874 \\
		        Wald test - $M = 10$ & 0.01 & 0.367 & 0 & 0.998 & 0.999 & 0.989 \\
		        \hline				        
	\end{tabular}
	\caption{Results for type I error comparing sub-samples from within the three cancer stages and power for comparison between the three pairs of cancer stages from the \texttt{breat} data set. The results are based on $1000$ bootstrap samples. }
	\label{tab:breast}
\end{table}

\section{Conclusion}
Hypothesis tests for covariance matrices in high dimension are challenging. RP based tests are known to be very efficient for mean vector testing in high dimensions. In this paper, we have developed the random projection based tests for the covariance matrix for both one and two sample tests. Standard multivariate tests such as LRT for the one sample test and Box-$M$ and Wald test for the two sample hypothesis have been studied after random projection into lower-dimensional space. Inference is based on the average $p$-value of $K$ random projections, where the rejection region is determined by the empirical critical values simulated under the null hypothesis using fixed covariance matrices. Through theorems \ref{thm:sphericity} and \ref{thm:twosample}, we have shown that the null distributions can be generated using identity matrices for the fixed covariance matrices. Simulation results have shown that RP based methods control type I error rates and achieve very good power over a wide range of models, whereas high dimensional methods have very inflated type I error rates. For the RP based methods, increasing the projection dimension $M$ lowers the type I error and increases power. In our limited simulation study, we have observed that a dimension of $M = 15$ achieves very good results. We applied the test procedures to two different gene expression data sets with $p = 2000$ genes. The results show that RP based tests preserve type I error even in real data applications whereas the current existing test procedures have inflated type I error rates. An interesting observation in the \texttt{breast} data is that all the tests have consistently high type I error for Stage IIB breast cancer data. This could be an indication that there is potentially high levels of heterogeneity in the data that is not captured by the covariance matrix alone.

RP based methods are known to be computationally intensive - with the computational cost being linear in $K$ and $M$. Typically, $K = 1000$ is large enough to obtain consistent results. Efficient methods for generating random matrices and parallelization can reduce the computational cost significantly. In spite of involving a matrix decomposition step, orthogonal random matrix generation is efficient since the matrix being decomposed is of low dimension ($M \times M$) and the projected dimension $M$ is generally chosen to be smaller than the sample size. Parallelizing the computations for different random projections matrices can achieve a significant reduction in the overall computational time. To this effect, we have developed an R package \texttt{cramp}, which is available to download from \href{https://github.com/dnayyala/cramp}{https://github.com/dnayyala/cramp}. Through efficient parallelization, \texttt{cramp} achieves very good computation times. Table \ref{tab:runtimes} present the run times to calculate the average $p$-values of the two sample RP-based test statistics for different combinations of $n$, $p$ and $M$ based on $K = 10^3$ random projections. All computations were done on R (ver. 4.0.2) running on a 3.6 GHz AMD Ryzen7 1800X processor with 64 GB RAM, parallelized on 12 cores. The runtime increases very slow with respect to all three quantities, with the maximum time being $10.97$ seconds. 
\begin{table}[!ht]
	\centering
	\begin{tabular}{|c|*{3}{c}|*{3}{c}|*{3}{c}|}
		\hline
		& \multicolumn{3}{c|}{$n = 20$} & \multicolumn{3}{c|}{$n = 40$}  &  \multicolumn{3}{c|}{$n = 50$} \\
		$p \downarrow M \rightarrow$ &  5	&	10	&	15	&	5	&	10	&	15	&	5	&	10	&	15	\\
		\hline
		100	&	2.65	&	3.16	&	3.46	&	2.64	&	3.22	&	3.64	&	2.62	&	3.29	&	3.6	\\
		200	&	2.69	&	3.66	&	3.61	&	2.76	&	3.06	&	3.52	&	2.66	&	3.52	&	3.95	\\
		500	&	2.73	&	3.55	&	3.18	&	2.76	&	3.74	&	3.23	&	2.7	&	3.43	&	3.35	\\
		1000	&	2.78	&	3.6	&	3.95	&	2.77	&	3.15	&	3.63	&	2.69	&	3.36	&	3.61	\\
		2000	&	2.8	&	3.73	&	10.97	&	3.51	&	3.52	&	10.82	&	3.75	&	4.09	&	10.83	\\
		\hline
	\end{tabular}
	\caption{Computation times (in seconds) of the RP-based test statistics for different values of $n, p$ and $M$ based on $K = 10^3$ random projections. }
	\label{tab:runtimes}
\end{table}
		
\section*{Appendix} 
\appendix

\noindent {\bf \proofname{ of Theorem \ref{thm:sphericity}}}

The proof of Theorem \ref{thm:sphericity} is along the same lines as the proof of Theorem 2 in \cite{Srivastava2014}. To show that the distribution of $\overline{\pi}_U$ is independent of $\sigma$, define $\xvec^*_{m; i} = \mathcal{R}_m \xvec_i, i = 1, \ldots, n, m = 1, \ldots, M$ as the projection of the $i^{\rm th}$ observation using the $m^{\rm th}$ random projection matrix. Then we have
\begin{equation*}
{\rm var} \left(\xvec_{m;1}, \ldots, \xvec_{m; n} \right) = \mathcal{S}^*_m = \mathcal{R}_m \mathcal{S} \mathcal{R}_m^{\top},
\end{equation*}
where $\mathcal{S}$ and $\mathcal{S}_m^*$ are the sample covariance matrices of the original and projected observations respectively. From equation \eqref{eqn:johnnagao2}, the p-values based on $M$ $i.i.d.$ random projection matrices are 
\begin{equation*}
\pi_m = 1 - \chi^2_{\nu} \left( \frac{1}{k} \tr \left\{ \frac{\mathcal{S}_m^*}{ \tr \mathcal{S}_m^* / k} - \mathcal{I}_k \right\}^2 \right).
\end{equation*}

Firstly since the random matrices are independent, conditional on the data $\mathcal{X} = \{\xvec_1, \ldots, \xvec_n\}$ and $\mathcal{Y} = \{\yvec_1, \ldots, \yvec_m  \}$, the p-values $\pi_1, \ldots, \pi_M$ are independent and identically distributed. This is because of the orthogonality of the projection matrices which preserves the covariance matrix structure ($\mathcal{R} \left(\sigma^2 \mathcal{I}_p \right) \mathcal{R}^{\top} = \sigma^2 \mathcal{I}_k$). Additionally, we can write
\begin{equation}
	P \left[ \overline{\pi} < u \right] = \mathbb{E}_{\mathcal{X}, \mathcal{Y}} \left\{ P_{\mathcal{R}} \left[ \overline{\pi} < u | \mathcal{X}, \mathcal{Y} \right]  \right\},
	\label{eqnapp:condexp}
\end{equation}
where the expected value is with respect to the distribution of the observations and the probability is with respect to the randomness of the projection matrix. 

By the conditional independence of $\pi_1, \ldots, \pi_M$ and the central limit theorem, we have a normal approximation to the probability in \eqref{eqnapp:condexp}
\begin{equation}
	\lim\limits_{M \rightarrow \infty} \left| P \left[ \overline{\pi} < u \right] - \Phi \left( \frac{u - \mathbb{E}_{\mathcal{R}} \left[u | \mathcal{X}, \mathcal{Y} \right]}{ {\rm var}_{\mathcal{R}} \left[ u | \mathcal{X}, \mathcal{Y} \right] } \right) \right| = 0.
	\label{eqnapp:clt}
\end{equation}
Hence the probability $P \left[ \overline{\pi} < u \right]$ can be approximated only using the moments of $U | \mathcal{X}, \mathcal{Y}$. Under the null hypothesis $H_{0S}$, the variable $U | \mathcal{X}, \mathcal{Y}$ is defined as
\begin{align}
U | \mathcal{X}, \mathcal{Y} & = 1 - \chi^2_{\nu} \left( U | \mathcal{X}, \mathcal{Y} \right) = 1 - F_{\chi^2_{\nu}} \left( \tr \left\{ \frac{\mathcal{S}_m^*}{ \tr \mathcal{S}_m^* / k} - \mathcal{I}_k \right\}^2 | \mathcal{X}, \mathcal{Y} \right) \nonumber \\
& \sim {\rm Unif}(0,1).
\label{eqnapp:unif}
\end{align}
The uniform distribution is from the standard property of $p$-value under the null hypothesis, which is independent of $\sigma^2$. Using this property, we shall show that the distribution of $E_{\mathcal{R}} \left[U | \mathcal{X}, \mathcal{Y} \right]$ and ${\rm var}_{\mathcal{R}} \left[U | \mathcal{X}, \mathcal{Y} \right]$ with respect to $\mathcal{X}, \mathcal{Y}$ are also independent of $\sigma^2$.

Let $W$ denote the expected value of $U | \mathcal{X}, \mathcal{Y}$ with respect to $\mathcal{R}$,
\begin{align}
	W = \mathbb{E}_{\mathcal{R}} \left[ U | \mathcal{X}, \mathcal{Y} \right] & = \int u \, dP_{\mathcal{R}} \nonumber \\
	& = \int \left[ 1 - F_{\chi^2_{\nu}} \left( \tr \left\{ \frac{\mathcal{S}_m^*}{ \tr \mathcal{S}_m^* / k} - \mathcal{I}_k \right\}^2 | \mathcal{X}, \mathcal{Y} \right)  \right] \, dP_{\mathcal{R}}
\end{align}
where the integral is with respect to the distribution of the random projection matrix $\mathcal{R}$. While the exact integral is not of importance, it should be noted that from equation \eqref{eqnapp:unif}, the integrand is independent of $\sigma^2$. As the random projection matrices are generated independent of the distribution of the observations, we can conclude that the variable $W$ is independent of $\sigma^2$. For any $m \geq 1$, the ${\rm m}^{\rm th}$ moment of $W$ is given by
\begin{align*}
	\mathbb{E}_{\mathcal{X}, \mathcal{Y}} \left[ W^m \right] = \int W^m \, dF_{\mathcal{X}, \mathcal{Y}} &= \int \mathbb{E}_{\mathcal{R}} \left[U| \mathcal{X}, \mathcal{Y} \right]^m  \, dF_{\mathcal{X}, \mathcal{Y}} \\
	& = \int \mathbb{E}_{\mathcal{R}} \left[U| \mathcal{X}, \mathcal{Y} \right] \times \cdots \times \mathbb{E}_{\mathcal{R}} \left[U| \mathcal{X}, \mathcal{Y} \right] \, dF_{\mathcal{X}, \mathcal{Y}}  \\
	& = \int \left\{ \int U_{\mathcal{R}_1} \, dP_{\mathcal{R}_1} \right\} \cdots \left\{ \int U_{\mathcal{R}_m} \, dP_{\mathcal{R}_m} \right\} \, dF_{\mathcal{X}, \mathcal{Y}}
\end{align*}
Interchanging the integrals by Fubini's theorem, we have
\begin{equation}
	\mathbb{E}_{\mathcal{X}, \mathcal{Y}} \left[ W^m \right]  = \int \cdots \int \left\{ \int U_{\mathcal{R}_1} \ldots U_{\mathcal{R}_m} \, dF_{\mathcal{X}, \mathcal{Y}} \right\} \, dP_{\mathcal{R}_1} \cdots dP_{\mathcal{R}_m}
\end{equation}
By the construction of $U$ in equation \eqref{eqnapp:unif}, the integral $\left\{ \int U_{\mathcal{R}_1} \ldots U_{\mathcal{R}_m} \, dF_{\mathcal{X}, \mathcal{Y}} \right\}$ is independent of $\sigma^2$. Therefore, all moments of $W$ are independent of $\sigma^2$ which implies that the distribution of $W$ is independent of $\sigma^2$. 

Similarly, it can be shown that the distribution of ${\rm var}_{\mathcal{R}} \left(U| \mathcal{X}, \mathcal{Y} \right)$ is also independent of $\sigma^2$. From the independence of the mean and variance, we have the distributions of
\begin{equation}
\Phi \left[ \frac{ u - \mathbb{E}_{\mathcal{R}} \left\{ U | \mathcal{X}, \mathcal{Y}  \right\}}{ {\rm var}_{\mathcal{R}} \left\{  U | \mathcal{X}, \mathcal{Y}  \right\} } \right]  \mbox{  and  }
\mathbb{E}_{\mathcal{X}, \mathcal{Y}} \left\{ \Phi \left[ \frac{ u - \mathbb{E}_{\mathcal{R}} \left\{ U | \mathcal{X}, \mathcal{Y}  \right\}}{ {\rm var}_{\mathcal{R}} \left\{  U | \mathcal{X}, \mathcal{Y}  \right\} } \right] \right\}
\label{eqnapp:indep}
\end{equation}
are independent of $\sigma^2$. Finally, combining this independence with equation \eqref{eqnapp:clt}, we have
\begin{gather*}
\lim\limits_{M \rightarrow \infty} P_\mathcal{R} \left\{ \overline{\pi} | \mathcal{X}, \mathcal{Y} \right\} = \Phi \left[ \frac{ u - \mathbb{E}_{\mathcal{R}} \left\{ U | \mathcal{X}, \mathcal{Y}  \right\}}{ {\rm var}_{\mathcal{R}} \left\{  U | \mathcal{X}, \mathcal{Y}  \right\} } \right] ,
\end{gather*}
with the right hand side independent of $\sigma^2$. Taking expected values with respect to $\mathcal{X}$ and $\mathcal{Y}$, we have
\begin{equation}
\lim\limits_{M \rightarrow \infty} P \left[ \overline{\pi} < u \right] = \mathbb{E}_{\mathcal{X}, \mathcal{Y}} \left\{ \Phi \left[ \frac{ u - \mathbb{E}_{\mathcal{R}} \left\{ U | \mathcal{X}, \mathcal{Y}  \right\}}{ {\rm var}_{\mathcal{R}} \left\{  U | \mathcal{X}, \mathcal{Y}  \right\} } \right] \right\}.
\label{eqnapp:limit}
\end{equation}
By equation \eqref{eqnapp:indep}, the right hand side in \eqref{eqnapp:limit} is also independent of $\sigma^2$, completing the proof. 

\qed

\noindent  {\bf \proofname{ of Theorem \ref{thm:twosample}}}

Invariance of the distribution of the two-sample test statistic can be shown similar to the above proof.  Besides computation of the test statistic, rest of the argument remains the same since the Box $M$ test statistic also follows a standard uniform distribution under the null hypothesis. Hence in Algorithm \ref{alg:algorithm2}, $\pi_m \sim {\rm Unif}(0, 1)$ under $H_0$, which is independent of the choice of $\Sigma$. 

\qed

\bibliographystyle{natbib}
\bibliography{./cramp_references_ver2}

\end{document}